\documentclass[article]{aa}
\usepackage{natbib}
\usepackage{graphicx}
\usepackage{color}

\begin{document}

\newcommand{\carcsec}{$\mbox{.\hspace{-0.5ex}}^{\prime\prime}$}

\title{Simulating Ellerman bomb-like events}
 \author{S.~Danilovic}

   \institute{Max-Planck-Institut f\"ur Sonnensystemforschung, Justus-von-Liebig-Weg 3, 37077 G\"ottingen, Germany}

   \date{\today}

\abstract
{Ellerman bombs (EB) seem to be a part of a whole spectrum of phenomena that might have the same underlying physical mechanism, but with observed characteristics which pose a considerable challenge to models.
}
{The aim of this study is to investigate whether the proposed mechanism, applied to the circumstances of EBs, produce the observed characteristics.  
}
{For this, realistic 3D MHD simulations are used. Two different cases are presented: the quiet Sun and an active region. 
}
{Both runs confirm that EB-like brightenings coincide with hot and dense plasma which is in agreement with predictions of 1D and 2D modellings. The simulated EB-like phenomena assume the observed flame-like form which depends on the complexity of the ongoing reconnection and the viewing angle. At the layers sampled by  H$\alpha$-wings, near temperature minimum and below, the magnetic field topology seem to be always the same. The field lines there trace the base of the current sheet and the reconnected $\cap$-loops.
}
{The EB features are caused by reconnection of strong-field patches of opposite polarity in the regions where the surface flows are the strongest. The weakest cases among them can be reproduced quantitatively by the current simulations.
}


\maketitle

\section{Introduction}\label{sec:introduction}

Ellerman Bombs \citep[EBs;][]{Ellerman1917} are defined as transient brightenings visible in the extended wings of the H$\alpha$ line. Recently these features attracted a lot of attention mainly for two reasons.

Firstly, they pose a considerable challenge to models because of their specific characteristics. High resolution observations show that they appear in the shape of a flame that seems to be rooted in the intergranular lanes \citep{Hashimoto2010,Watanabe2011}. This suggests that their formation begins very low, near the surface. However, one- and two-dimensional modelling strongly suggests that only a temperature increase starting at heights of a few hundred km above the solar surface can produce the observed H$\alpha$ line profile without continuum brightening \citep{Kitai1983,Fang:2006,2006SoPh..235...75S,Bello:etal:2013,Berlicki:2014,Hong:2014,2016A&A...593A..32G}. EBs have signatures also in Ca~II~H and Ca~II~IR~854~nm \citep{Matsumoto2008,Gregal:2013,Reza2015}, but not in Na~I~D1 and Mg~I~b1 lines \citep{Rutten2015}. This indicates that temperatures are high enough so that neutral metals are ionized, but not as high so they loose their visiblity in chromospheric lines. In some cases, though, EBs can be traced also in observables which should sample temperatures orders of magnitude higher \citep{Vissers2015,Tian2016,Libbrecht2016}.

Secondly, they seem to be a part of a whole spectrum of phenomena that might have the same physical mechanism behind. On one hand, on the low energy end of that spectrum are quiet-Sun Ellerman-like brightenings \citep[QSEBs;][]{Luc2016}. As EBs, these features also produce emission in H$\alpha$ wings, but they are less bright and smaller in size. On the other hand, at the high-energy end of the spectrum sit so-called IRIS bombs \citep[IBs;][]{Peter2014} which in some cases coincide with EBs \citep{Tian2016}, but seem to point to much higher temperatures. The specific signatures like presence of absorption
lines in greatly broadened profiles of transition region lines indicates local
heating in the photosphere or lower chromosphere to $ \approx 8\times10^{4}$~K.

Observations show that EBs are preferably formed in young emerging active regions \citep{Bruzek1968,Rutten2013,Schmieder2014}. They usually, but not exclusively, appear in series where the emergence of serpentine field lines takes place, in so-called Bald Patches (BPs) - dips in magnetic field lines \citep{Pariat:etal:2004,Pariat:2006,2009ApJ...701.1911P,2012ASPC..455..177P}. This scenario was supported by the ideal MHD simulations \citep{Isobe2007,Archontis2009}, who generated the temperature and density increase of the order of magnitude that was needed to reproduce the observed H$\alpha$ wings, but did not allow any quantitative comparison with observations due to their simplified treatment of the plasma physics. 


In \cite{ja:sunrise2}, we quantitatively compared our 3D realistic MHD simulations with high-resolution observations and showed, for the first time, that serpentine-like emergence indeed produces EB-like phenomena. There, we focused on the dynamics of simulated EB-like events and their signatures in photospheric neutral iron lines. The current paper further explains what causes their flame like morphology and shows similarities between the quiet Sun and active region case. 

\begin{figure*}
  \centerline{\includegraphics[angle=-90,%
     width=\linewidth,trim= 0cm 0cm 0cm 0cm,clip=true]{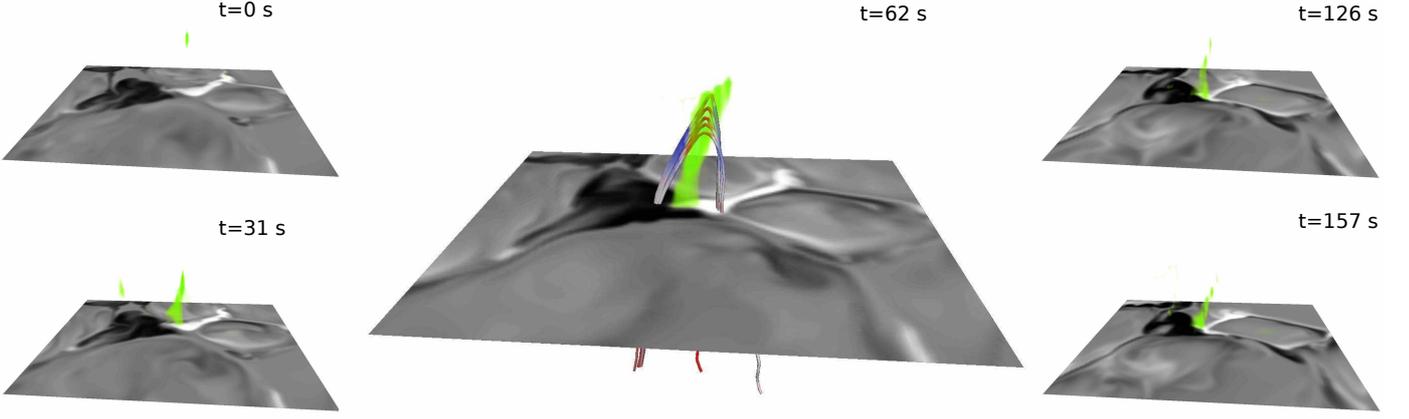}}
     \caption[]{\label{fig:muram1-3d} %
     Results from the weak-field MURaM simulation at half-minute
     intervals, ordered along columns.
     The grey scale in the horizontal planes shows the vertical
     component of the magnetic field at the geometrical height where $\langle \tau \rangle = 0.1$.
     The third sample is enlarged for better visibility.
     Green denotes volume rendering of temperature in the range $T=5700-6500$~K and outlines the site with most Ohmic heating.
     Some magnetic field lines are shown in the middle panel,
     color-coded according to vertical gas velocity (downflows red,
     upflows blue).
     The sequence illustrates that cancellation of opposite-polarity
     field concentrations produces heated features with EB-like
     behaviour.
     }
\end{figure*}

\begin{figure}
  \includegraphics[width=\columnwidth]{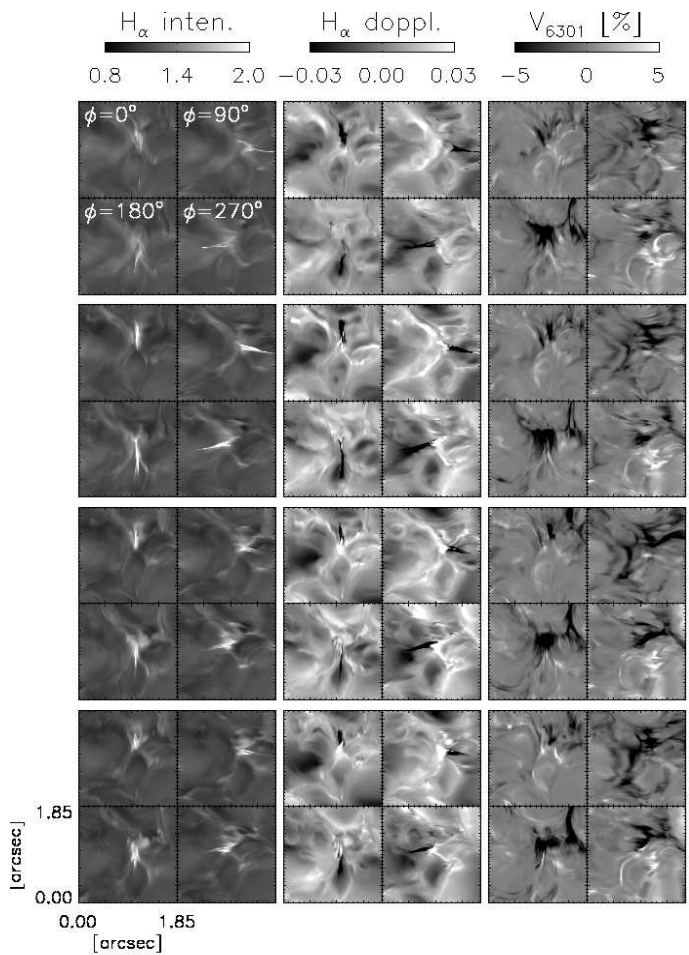}
     \caption[]{\label{fig:muram1-evolution} %
     Details of the EB-like feature for the last four
     samples in Fig.~\ref{fig:muram1-3d}, ordered top to bottom.
     For each sample time, the first column shows a quartet of
     synthetic intensity images in the H$_{\alpha}$ wing at
     $\Delta\lambda=-0.11$~nm from line center at viewing angle
     $\theta=49^{o}$, from four different azimuthal viewing directions
     $\phi$ as specified.
     The second column shows corresponding synthetic H$_{\alpha}$ 
     Dopplergrams obtained by subtracting blue-wing from red-wing
     images at $\Delta\lambda=\pm 0.11$~nm.
     The third column shows corresponding synthetic magnetograms
     obtained from the Fe~I~6301~\AA line at
     $\Delta\lambda=-48$~m\AA\ from line center.}
\end{figure}

\begin{figure}
  \includegraphics[width=0.45\columnwidth]{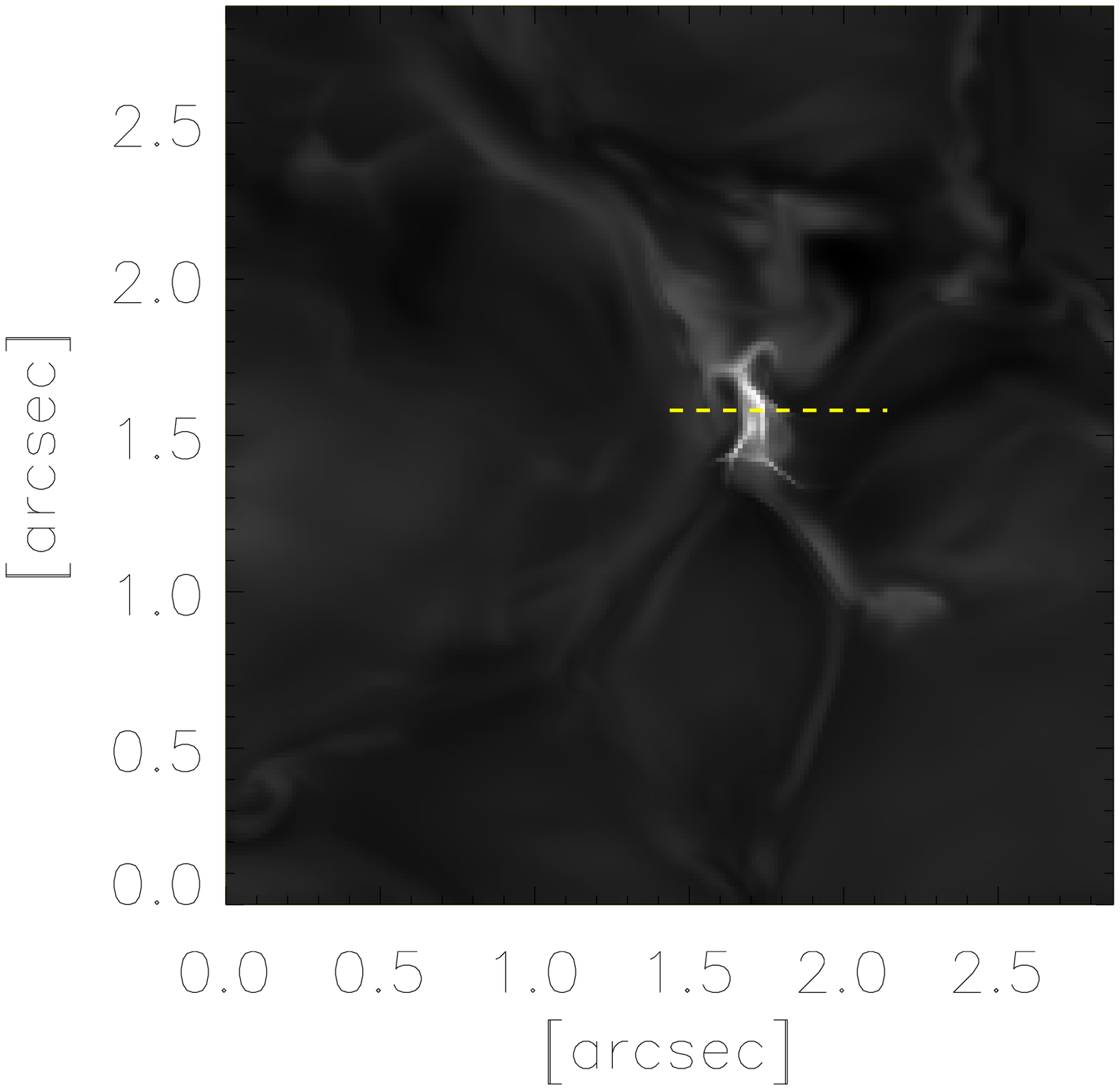}
  \includegraphics[width=0.45\columnwidth]{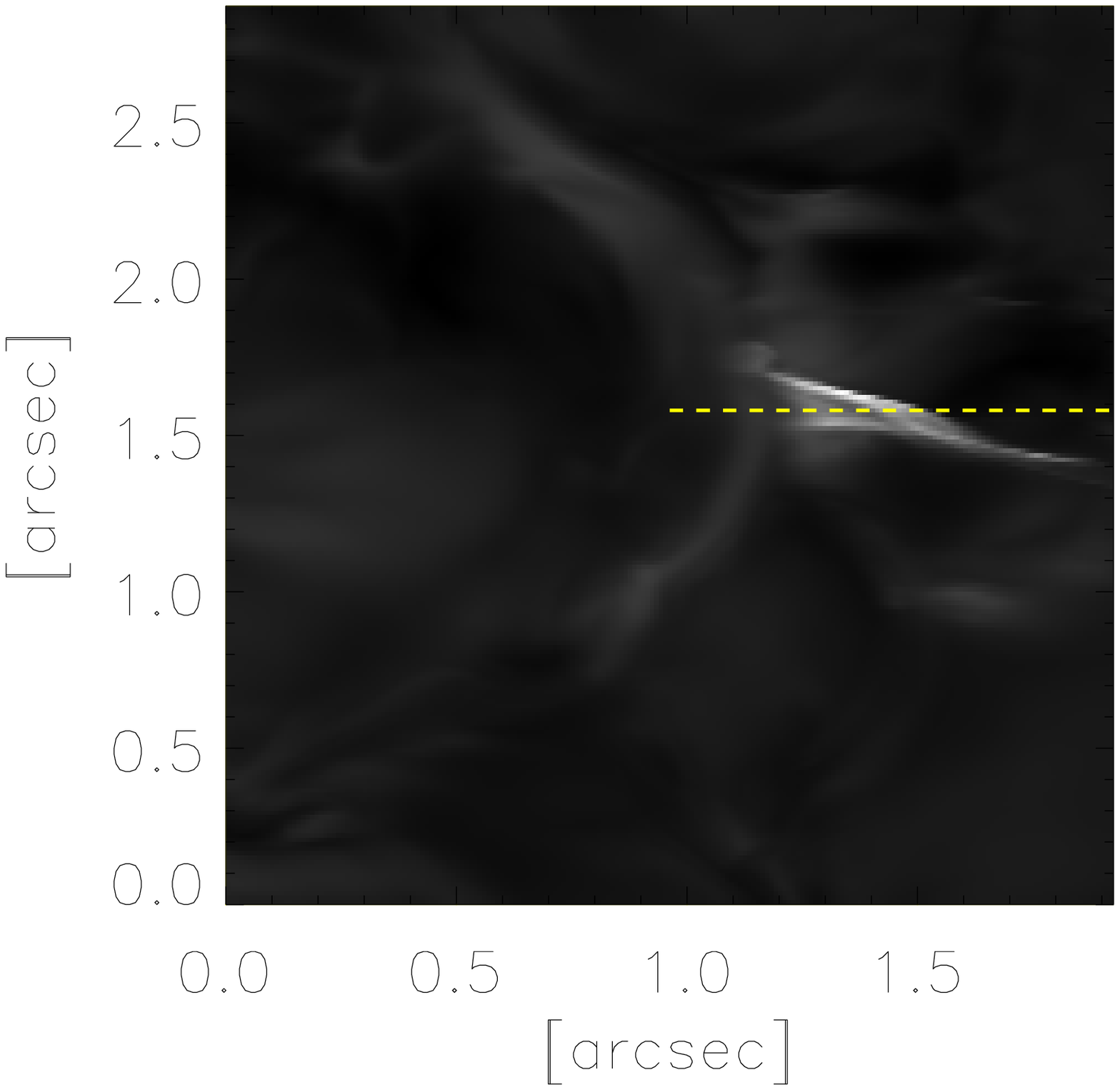} \\
  \includegraphics[width=0.45\columnwidth]{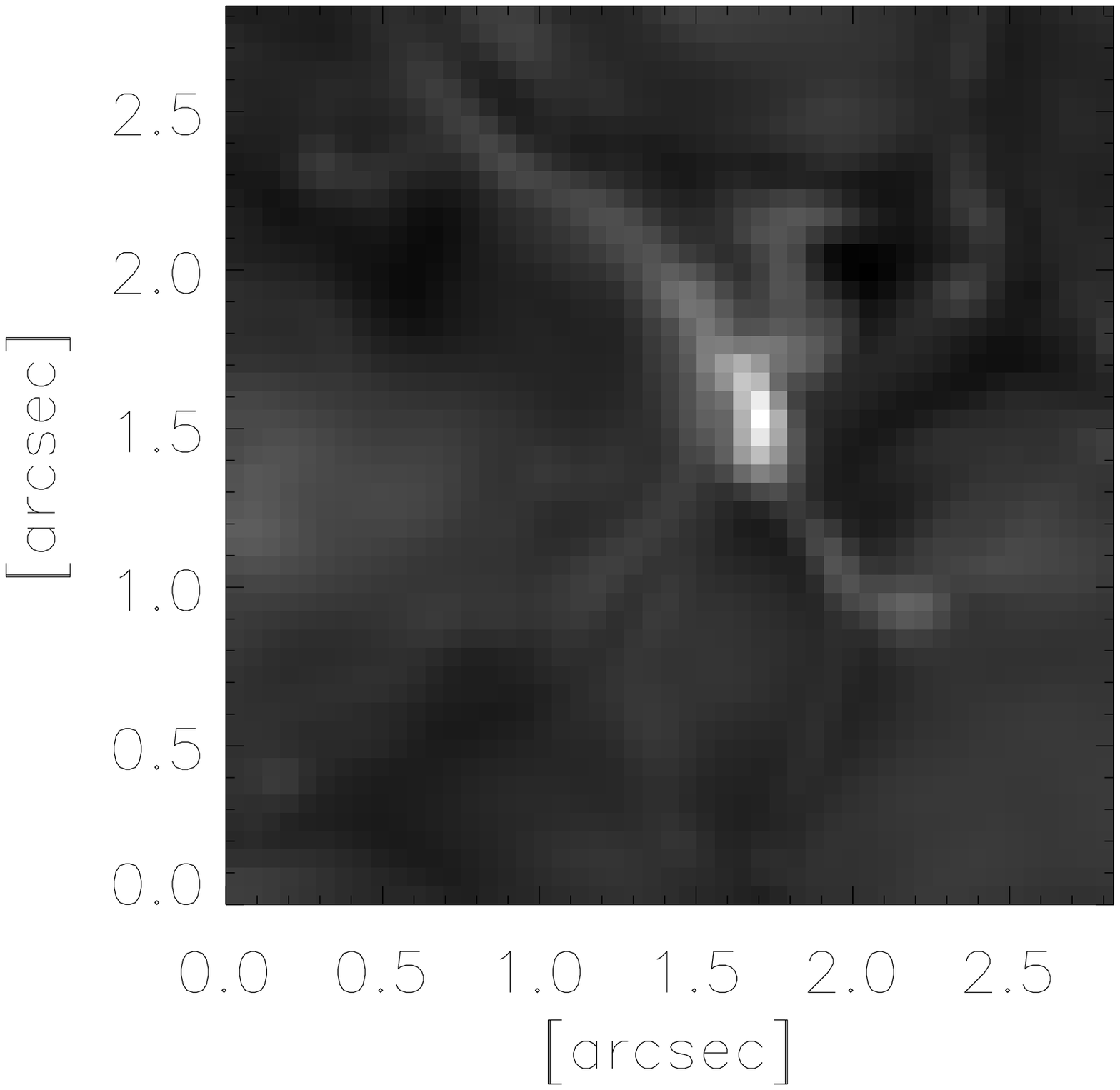}
  \includegraphics[width=0.45\columnwidth]{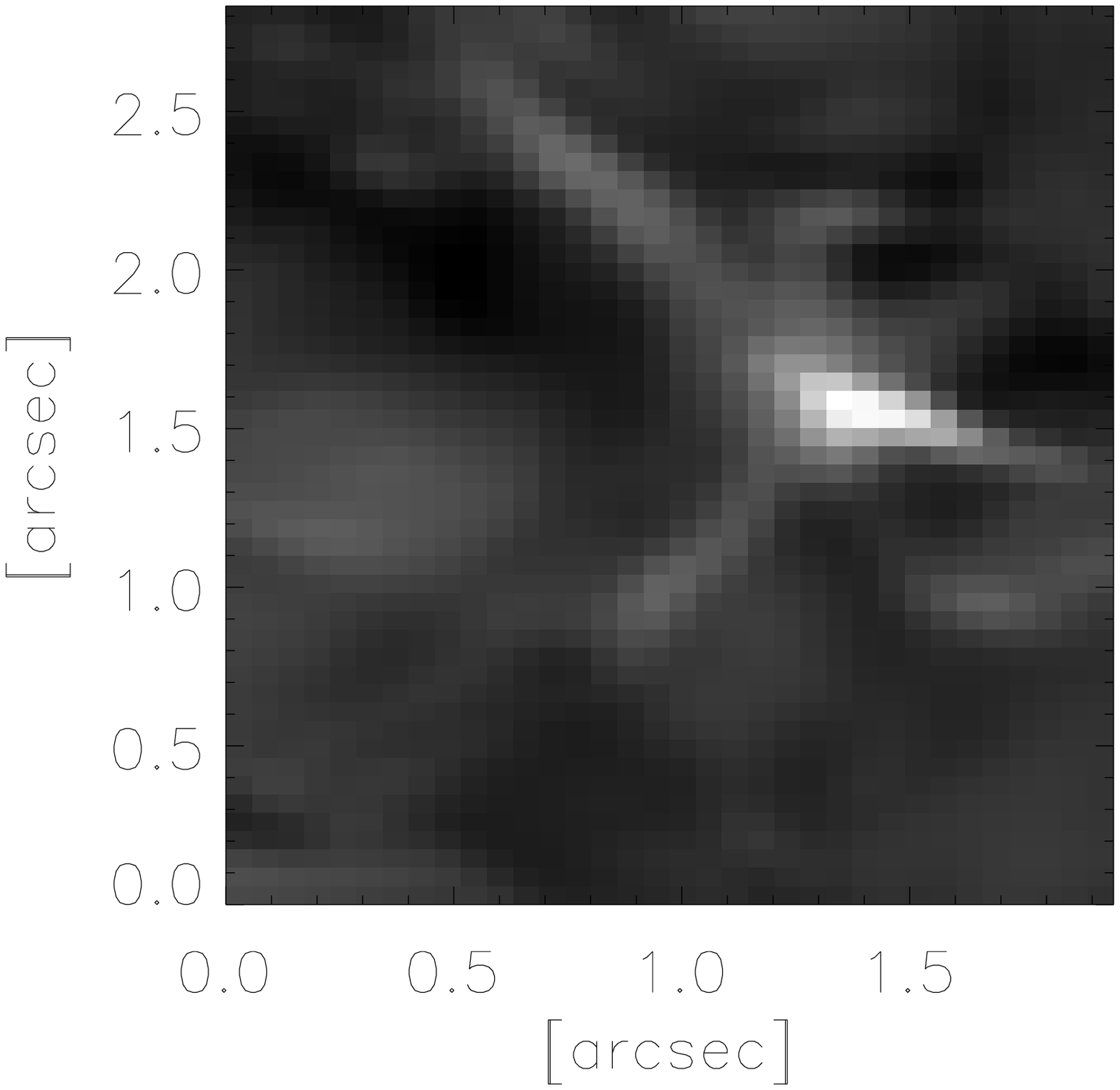} \\
  \includegraphics[width=0.45\columnwidth]{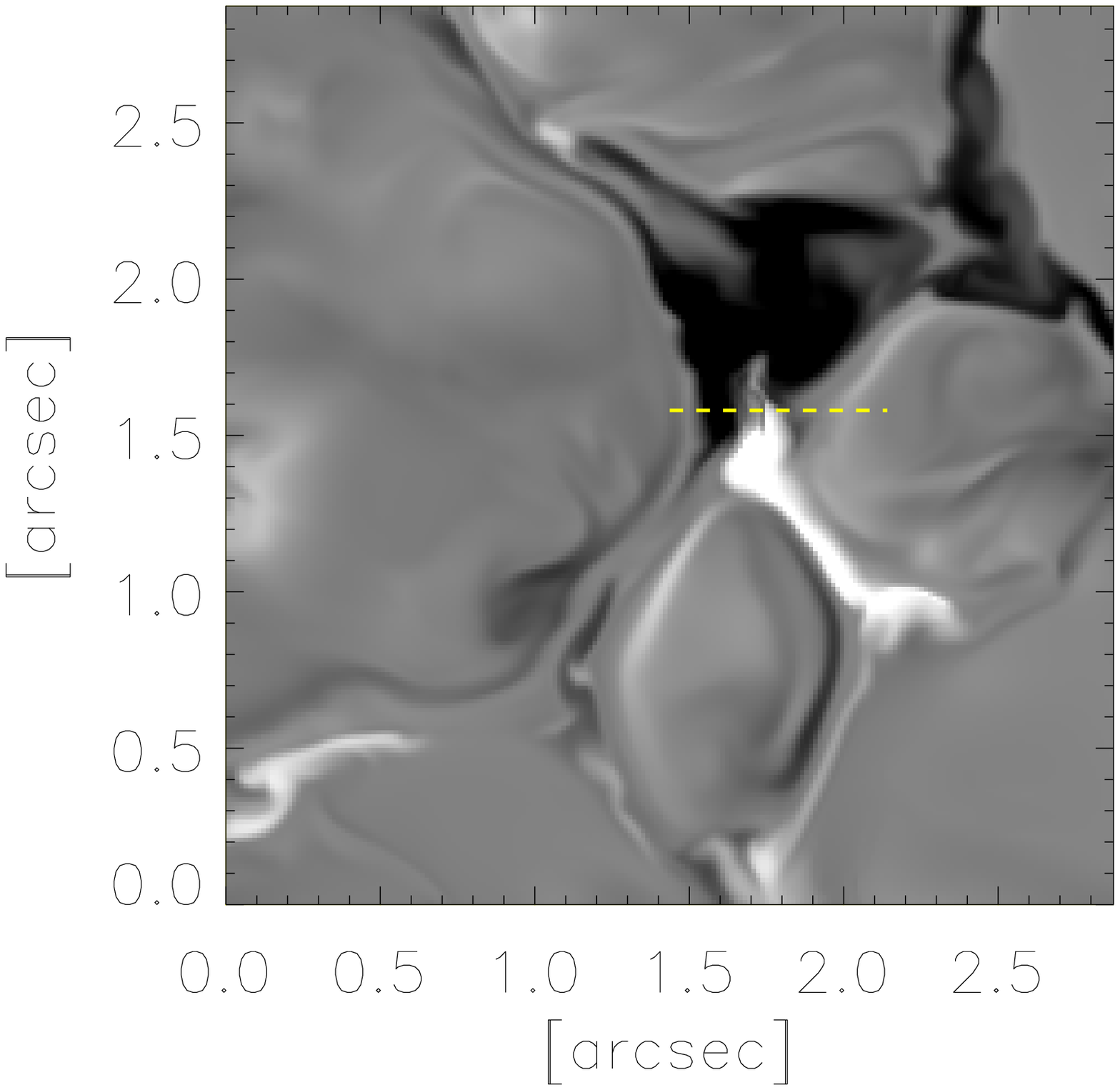}
  \includegraphics[width=0.45\columnwidth]{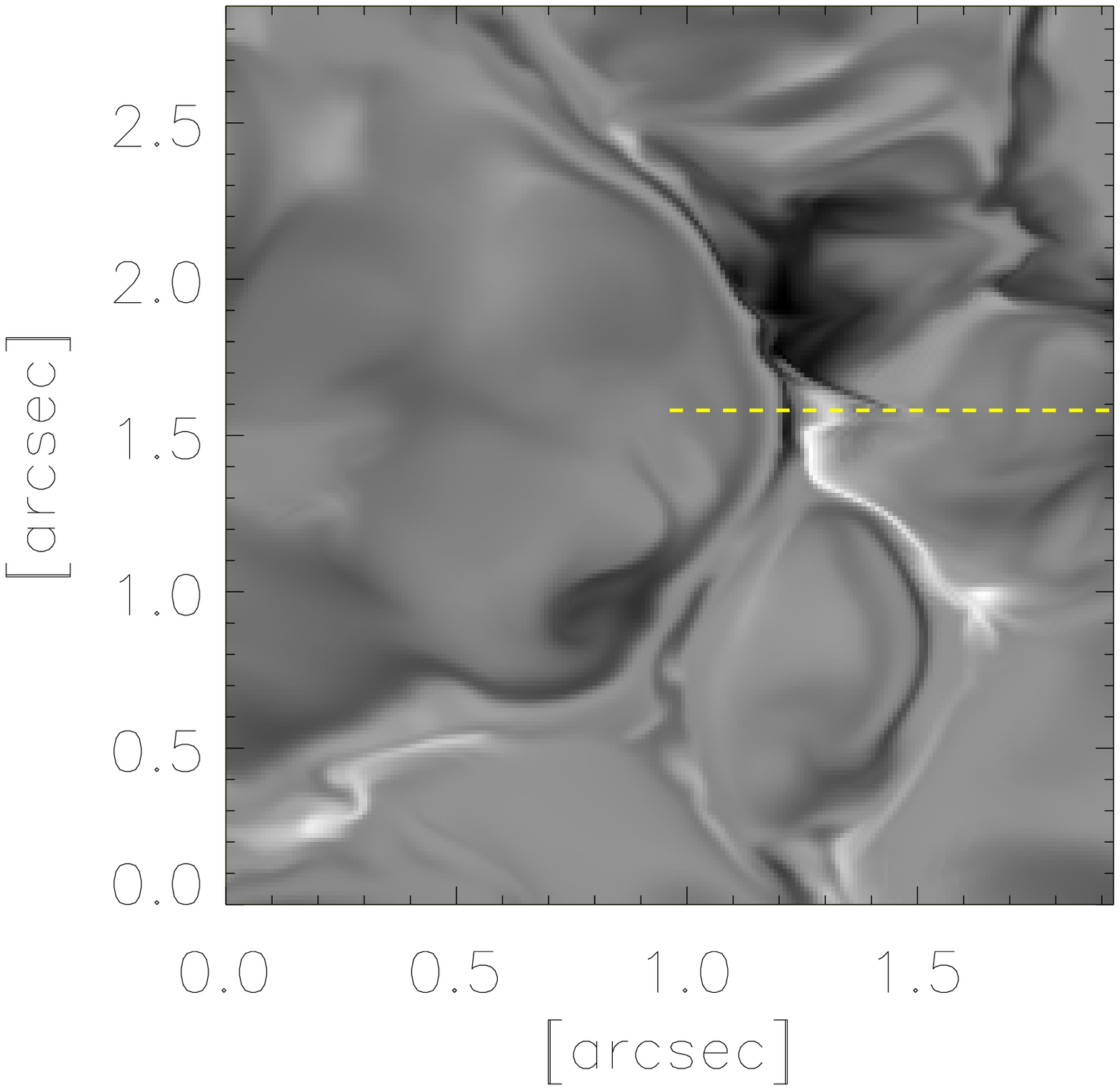} 
  \caption[]{\label{fig:muram1-cuts-map} %
  Synthesized images observed at disk center (first column) and at
  viewing angle $\theta=49^{o}$ and azimuthal angle $\phi=90^{o}$ (second column).
  Both columns are for time $t=62$~s (third sample in
  Fig.~\ref{fig:muram1-3d}, second row in
  Fig.~\ref{fig:muram1-evolution}).
  Top row: H$_{\alpha}$  wing intensity at simulation resolution.
  Middle row: H$_{\alpha}$  wing intensity at SST resolution.
  Bottom row: Fe~I~6301~\AA magnetograms at simulation resolution.
  The yellow dashed lines in the top and bottom panels mark the
  location of the vertical cuts shown in
  Figs.~\ref{fig:muram1-cuts-th0} and \ref{fig:muram1-cuts-th49}.
  }
\end{figure}

\begin{figure*}
  \includegraphics[angle=90,width=0.32\linewidth]{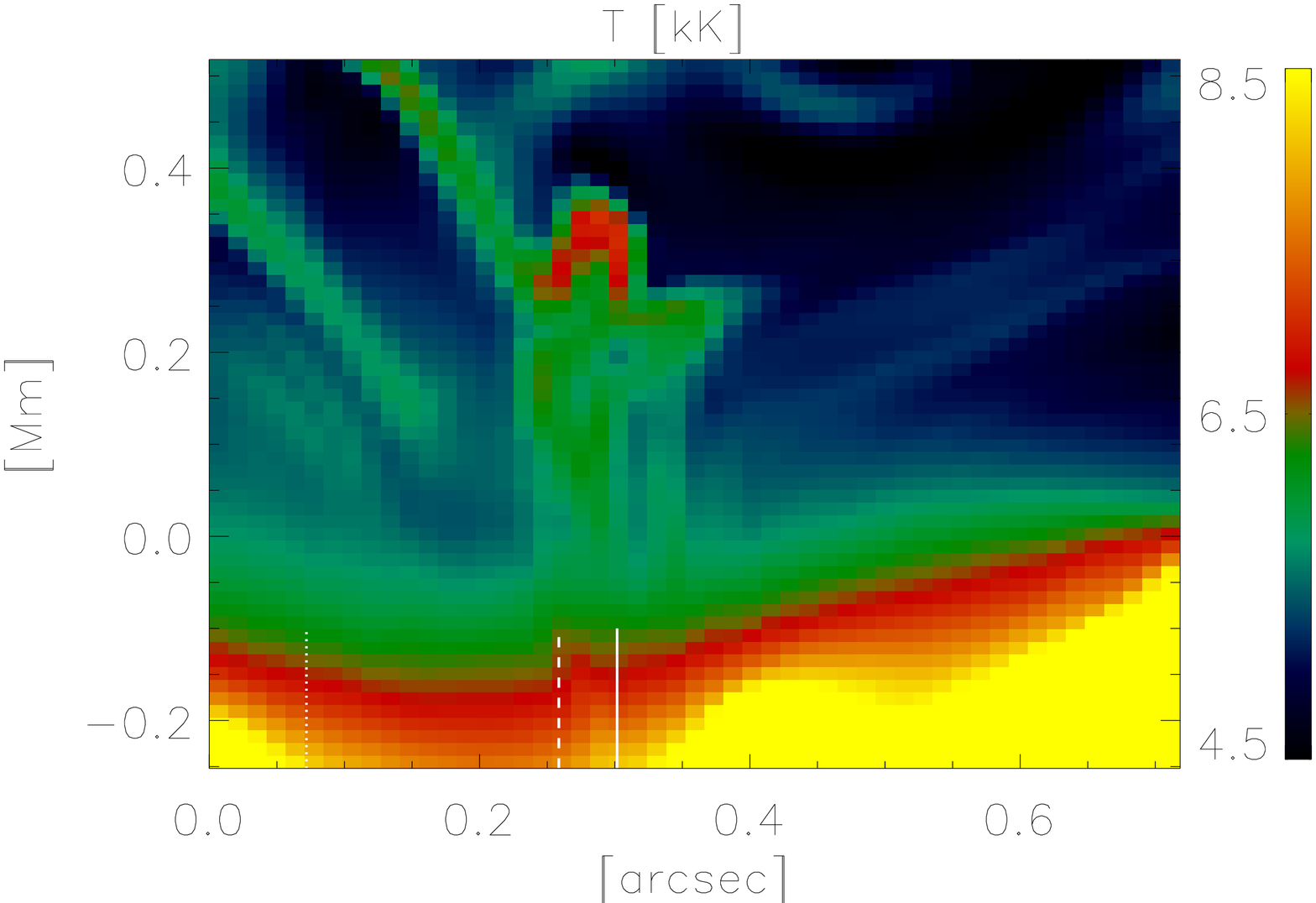}
  \includegraphics[angle=90,width=0.32\linewidth]{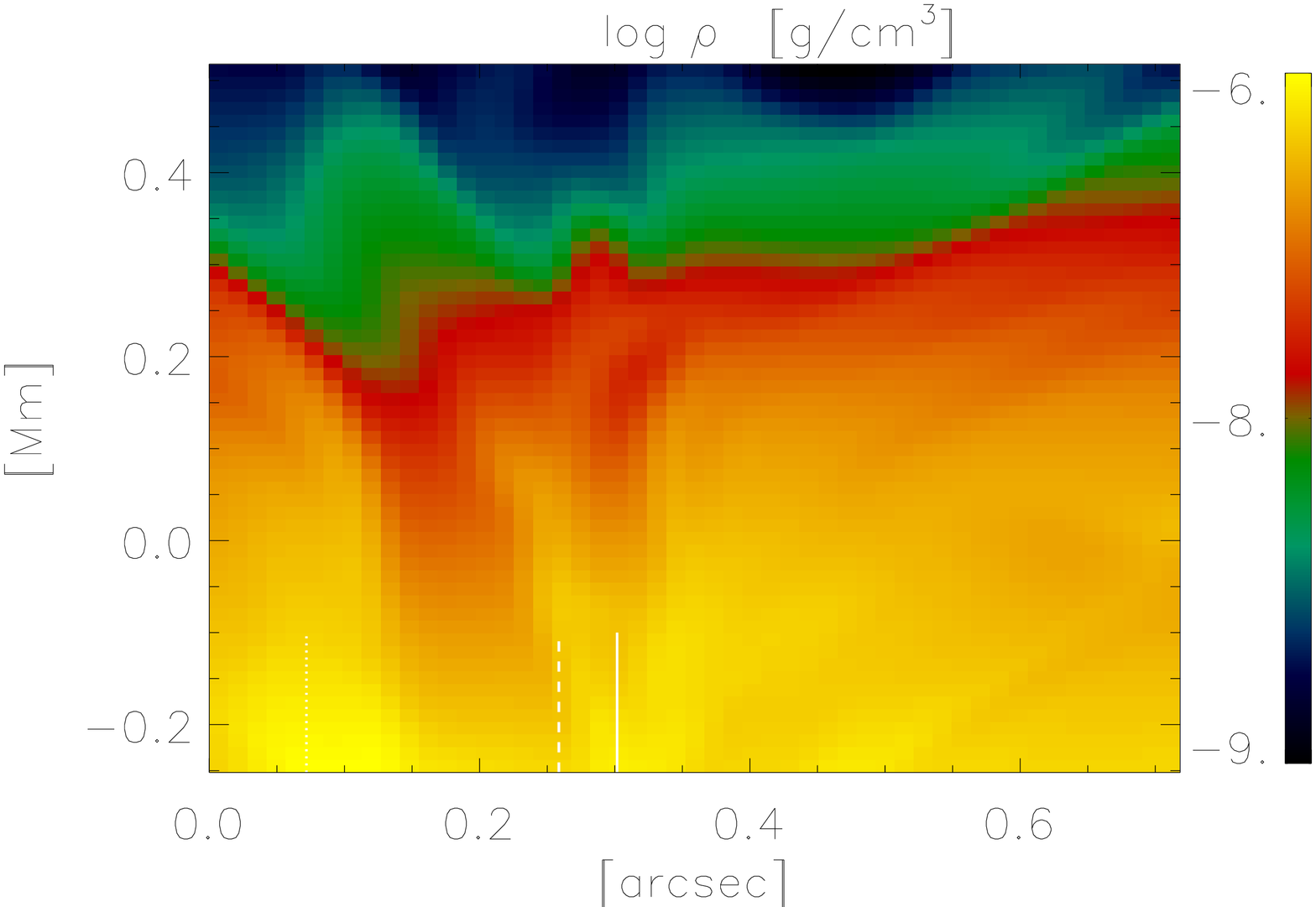}
   \includegraphics[angle=90,width=0.32\linewidth]{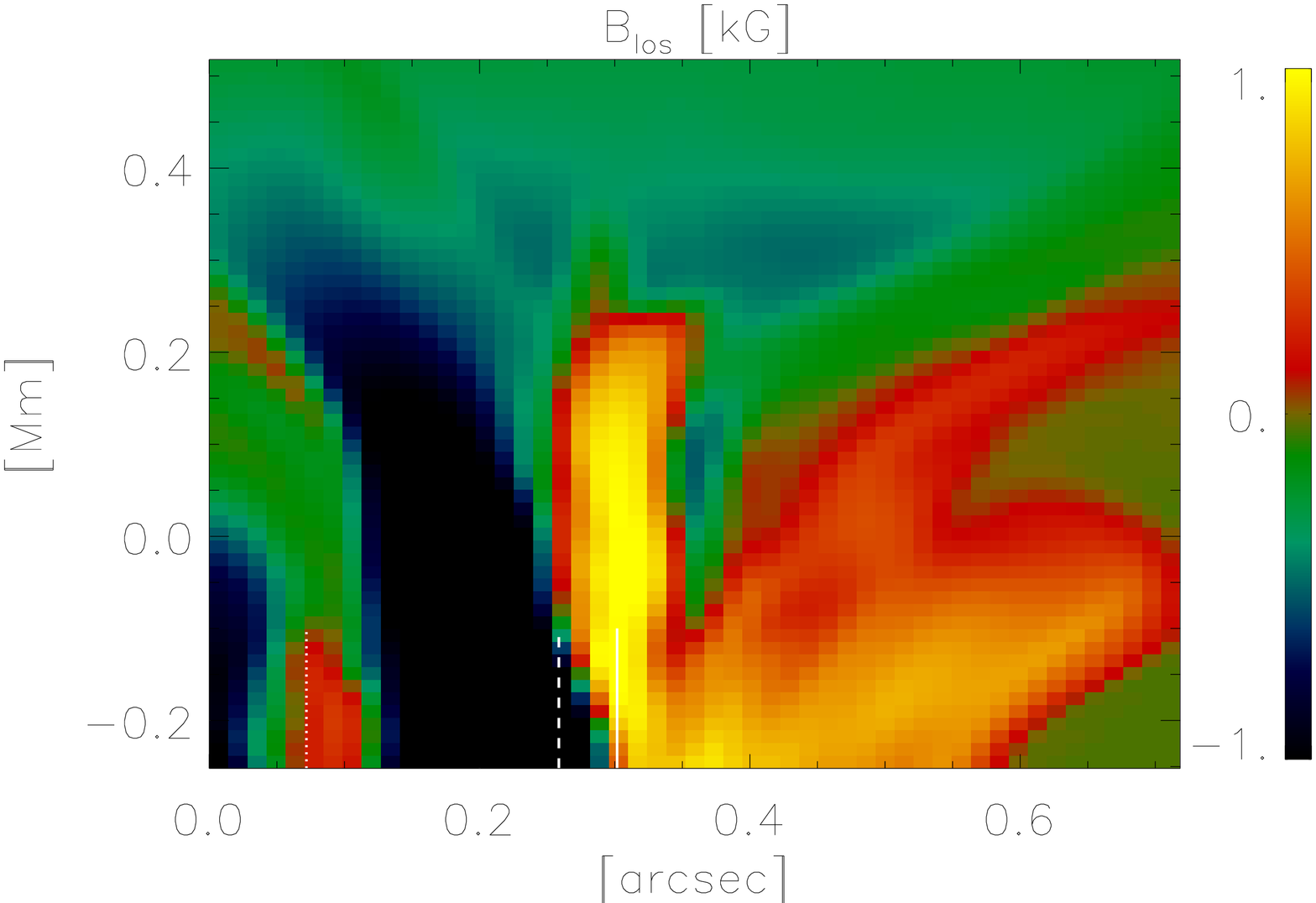} \\
  \includegraphics[angle=90,width=0.32\linewidth]{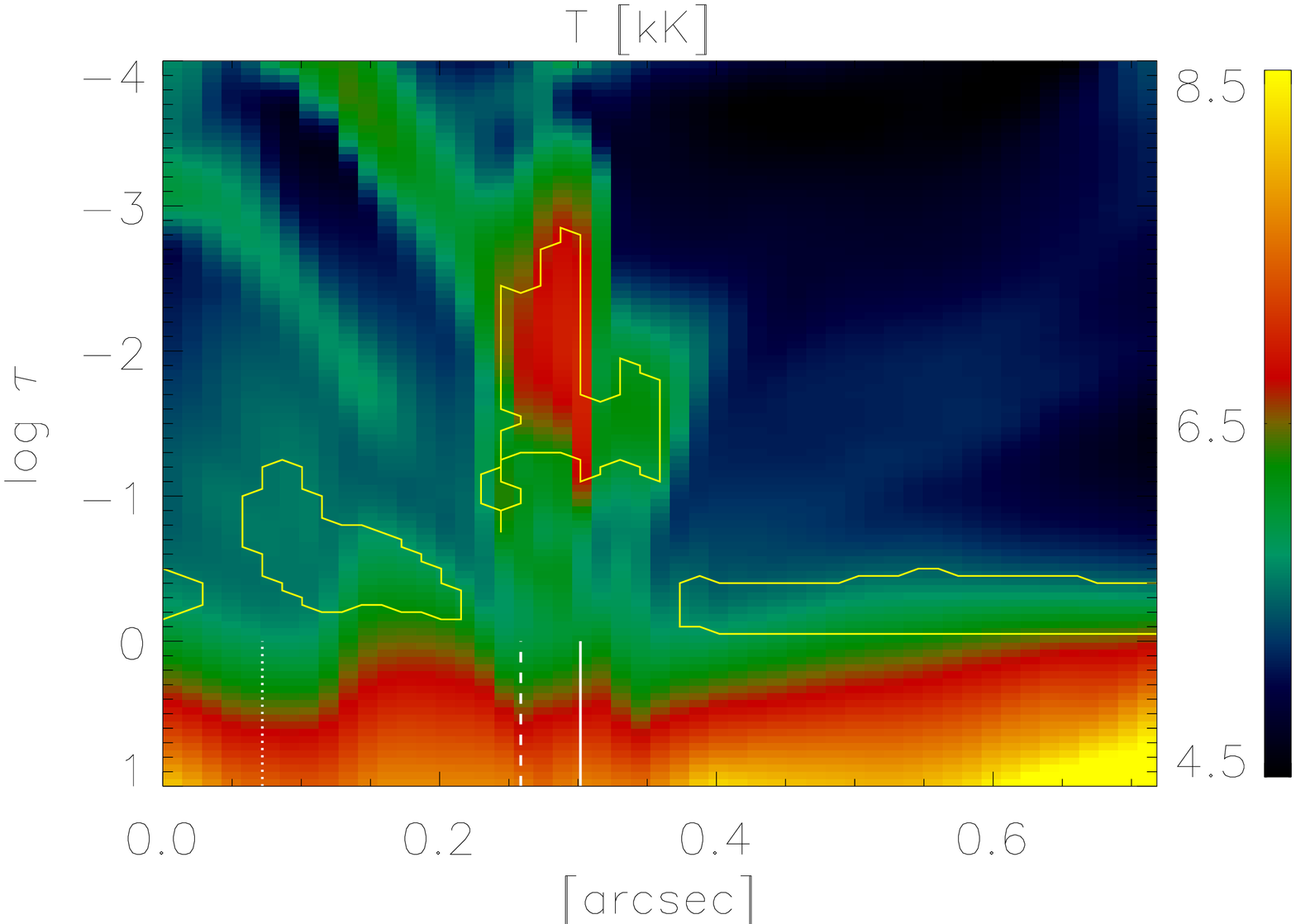}
  \includegraphics[angle=90,width=0.32\linewidth]{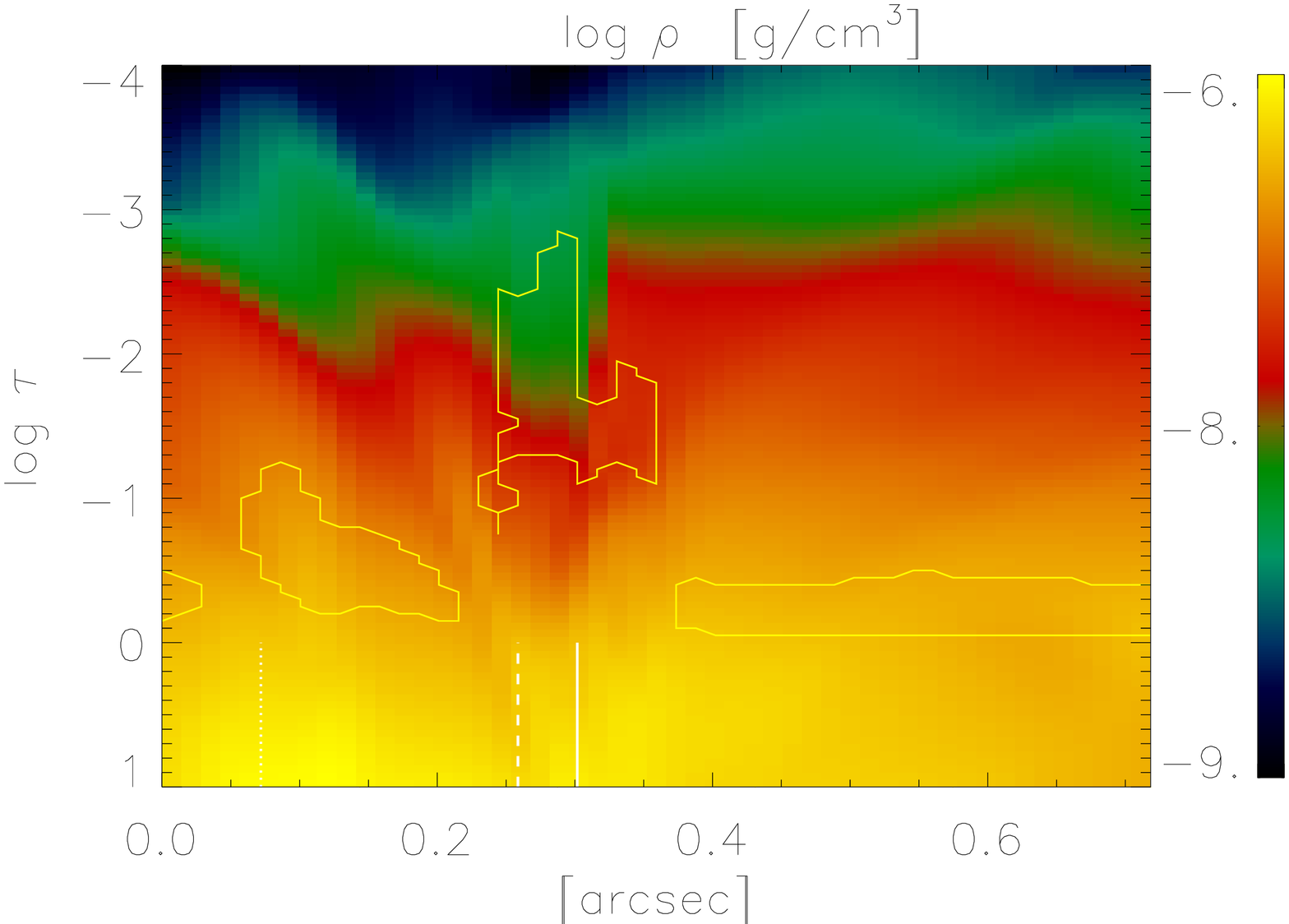}
   \includegraphics[angle=90,width=0.32\linewidth]{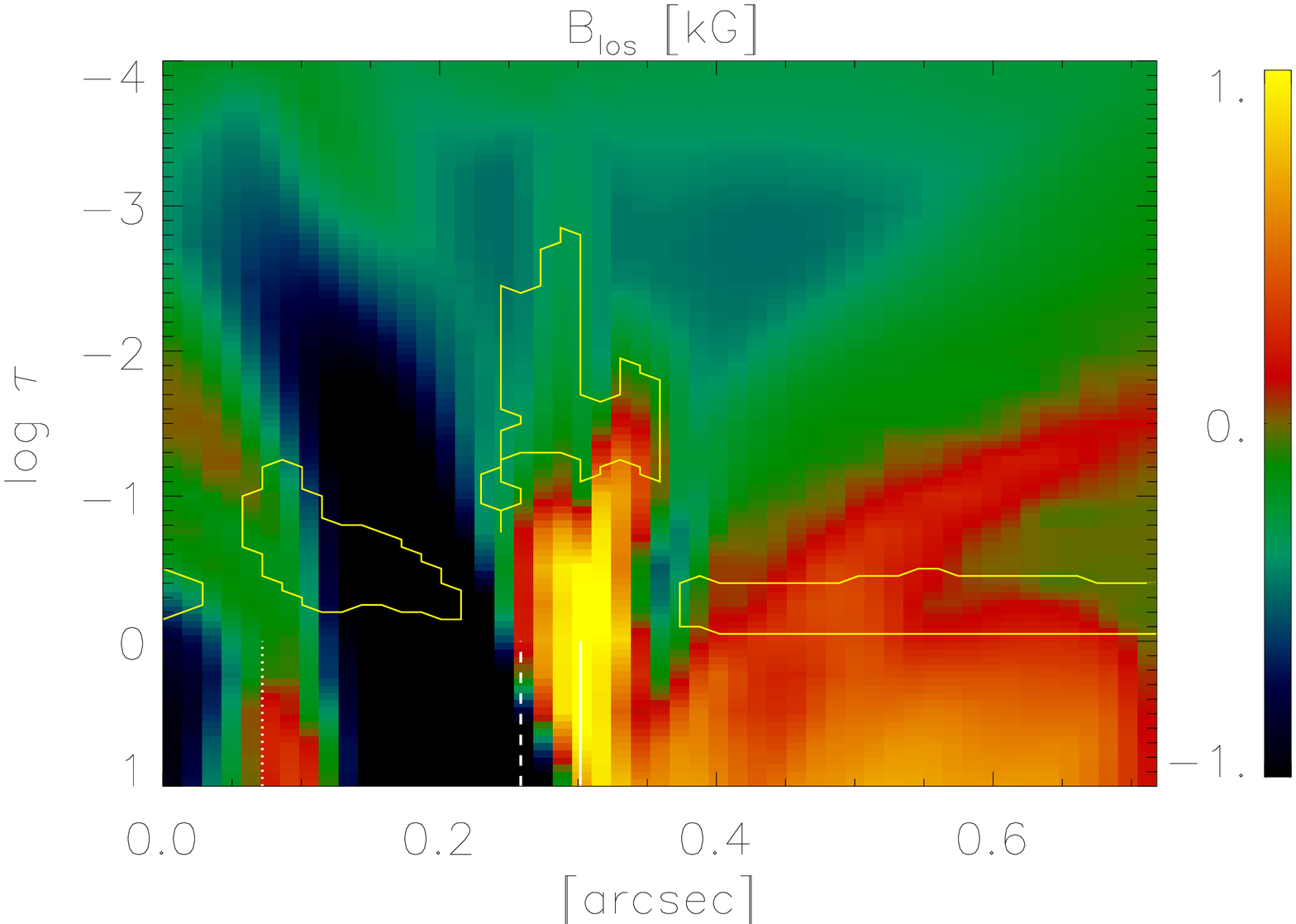} \\
  \includegraphics[angle=90,width=0.32\linewidth]{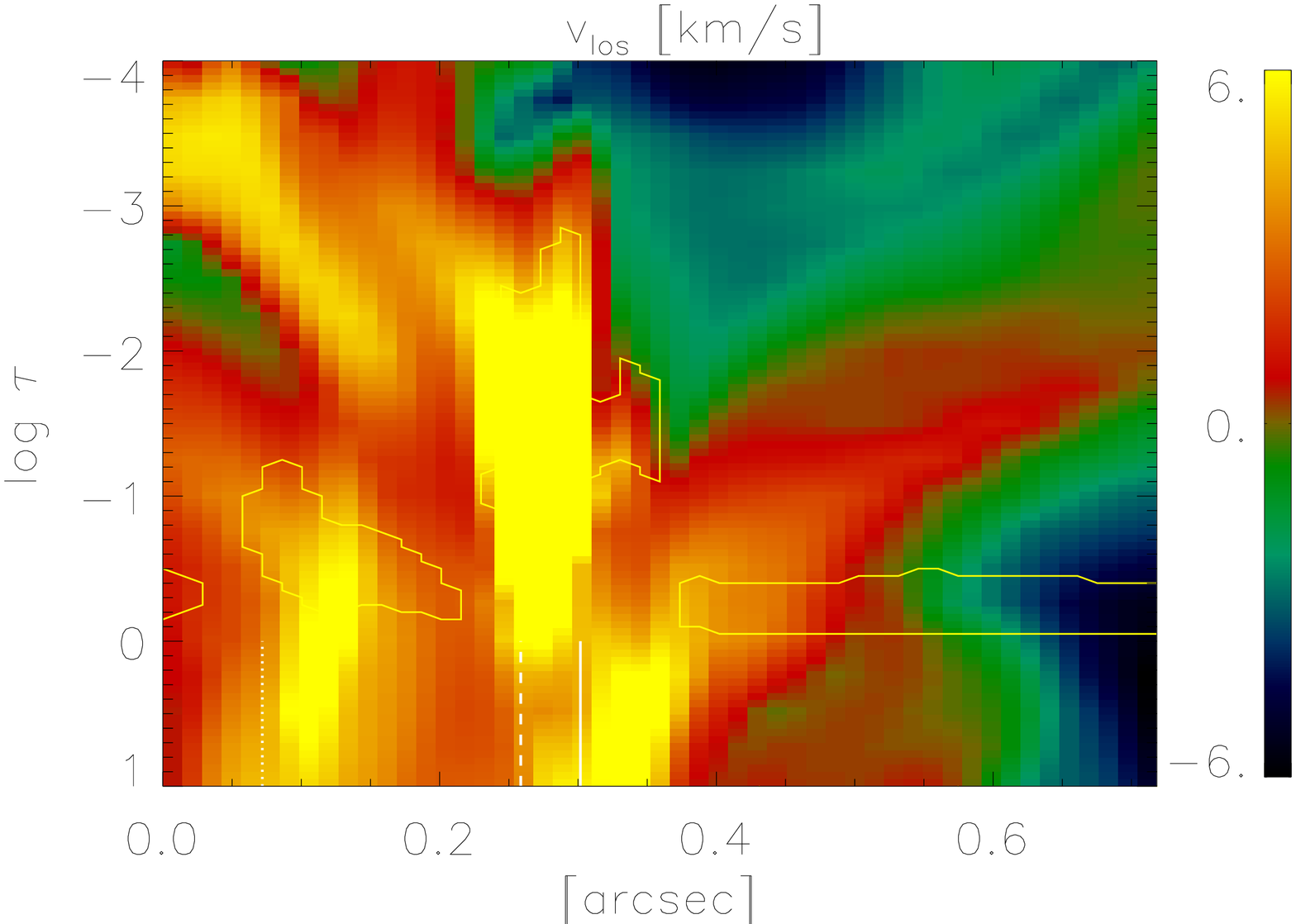} 
  \includegraphics[angle=90,width=0.32\linewidth]{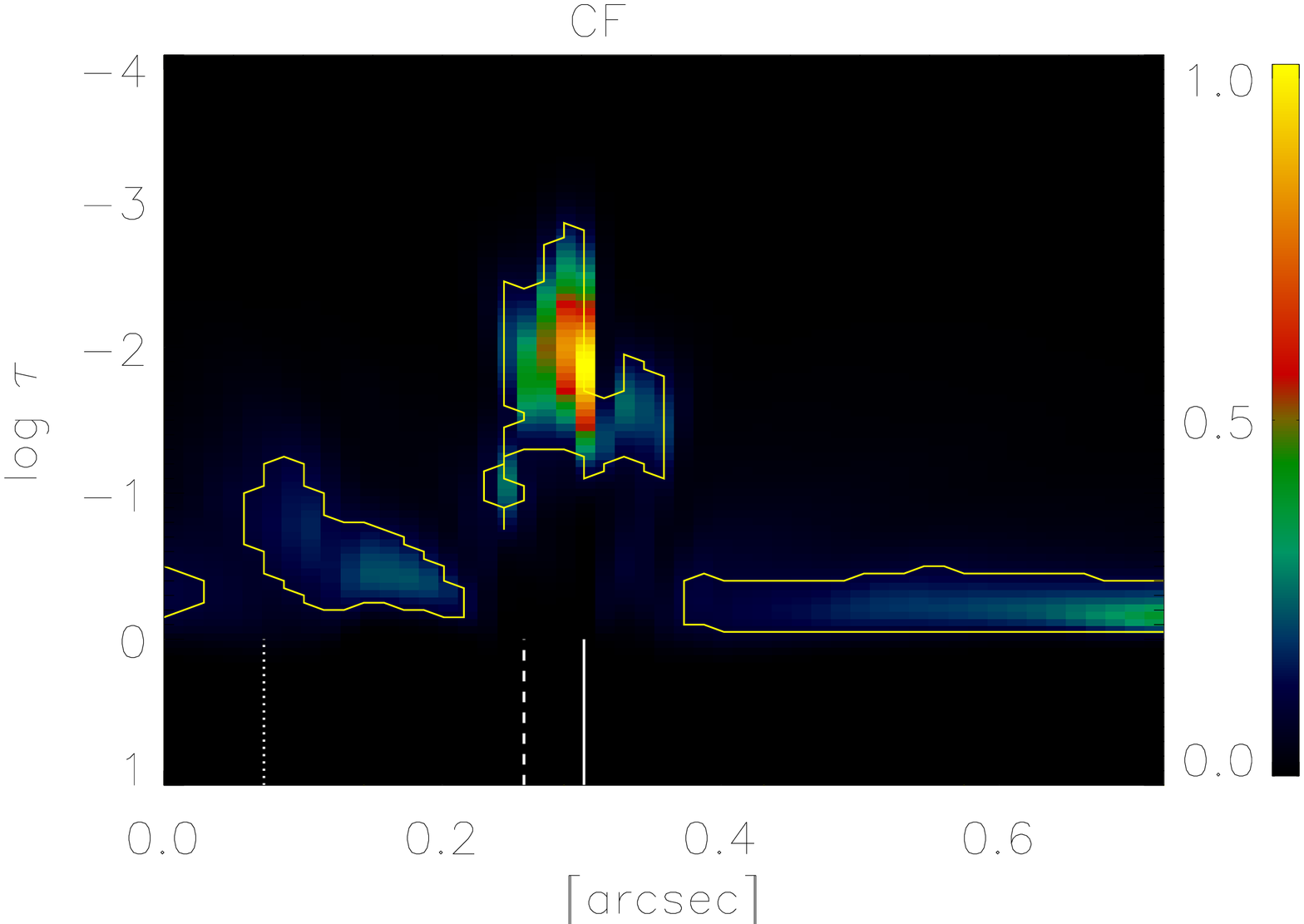} 
  \includegraphics[angle=90,width=0.32\linewidth]{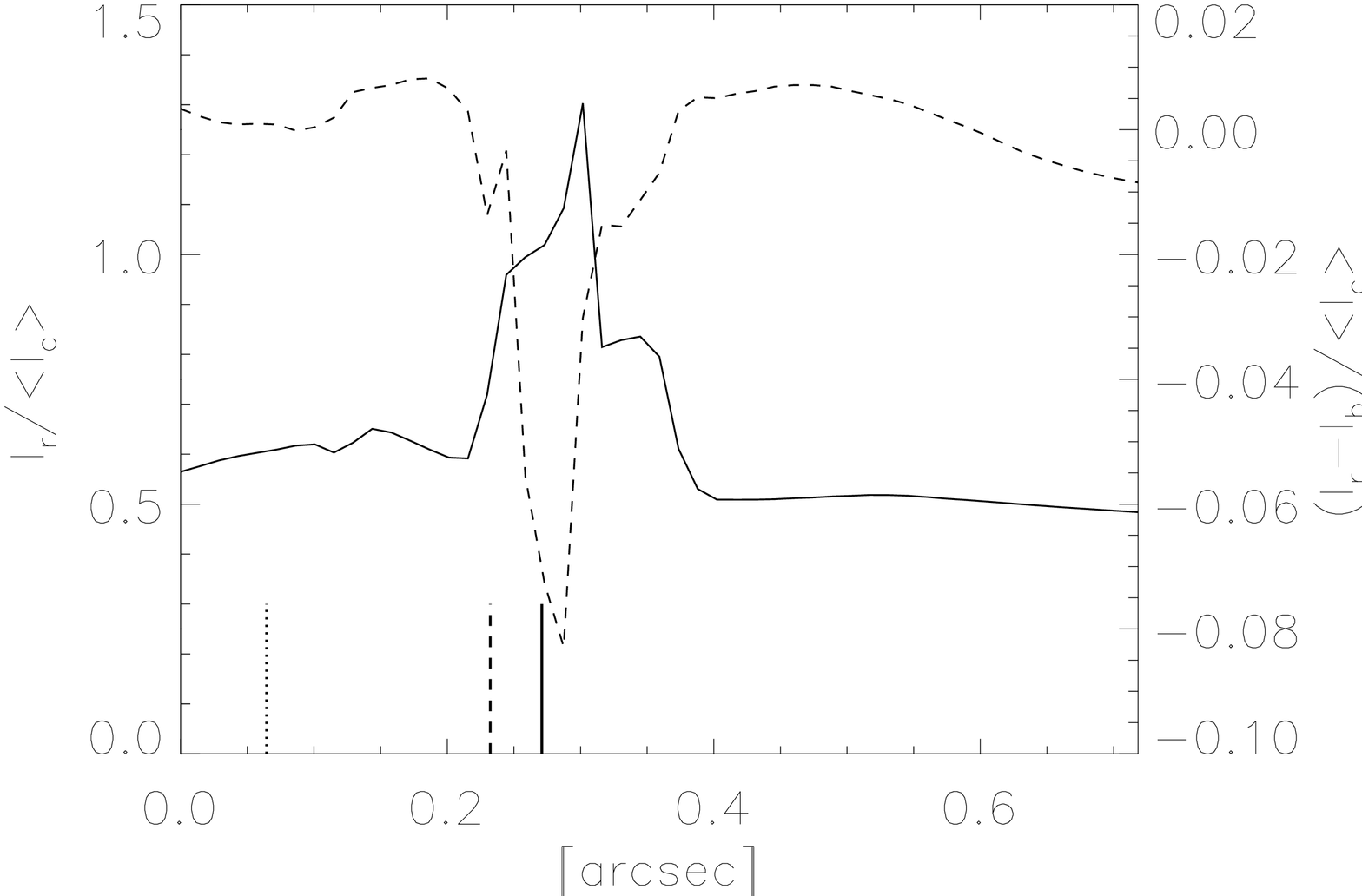} \\ 
  \caption[]{\label{fig:muram1-cuts-th0} %
  Vertical cuts for viewing angle $\theta = 0^{o}$ at the
  location shown in Fig.~\ref{fig:muram1-cuts-map}.
  \textit{First row\/}: temperature, density, line-of-sight component
  of the magnetic field against geometrical height.
  \textit{Second row\/}; temperature, density, line-of-sight field
  against optical depth.
  \textit{Third row\/}: line-of-sight velocity, contribution function
  to H$_{\alpha}$  line intensity at $\Delta\lambda=- 0.11$~nm from line
  center and H$_{\alpha}$ wing intensity overplotted with Dopplergram
  signal (dashed).
  The yellow contours outline the formation height of the blue
  H$_{\alpha}$  wing (central panel in the bottom row).
  }
\end{figure*}

\begin{figure*}
  \includegraphics[angle=90,width=0.32\linewidth]{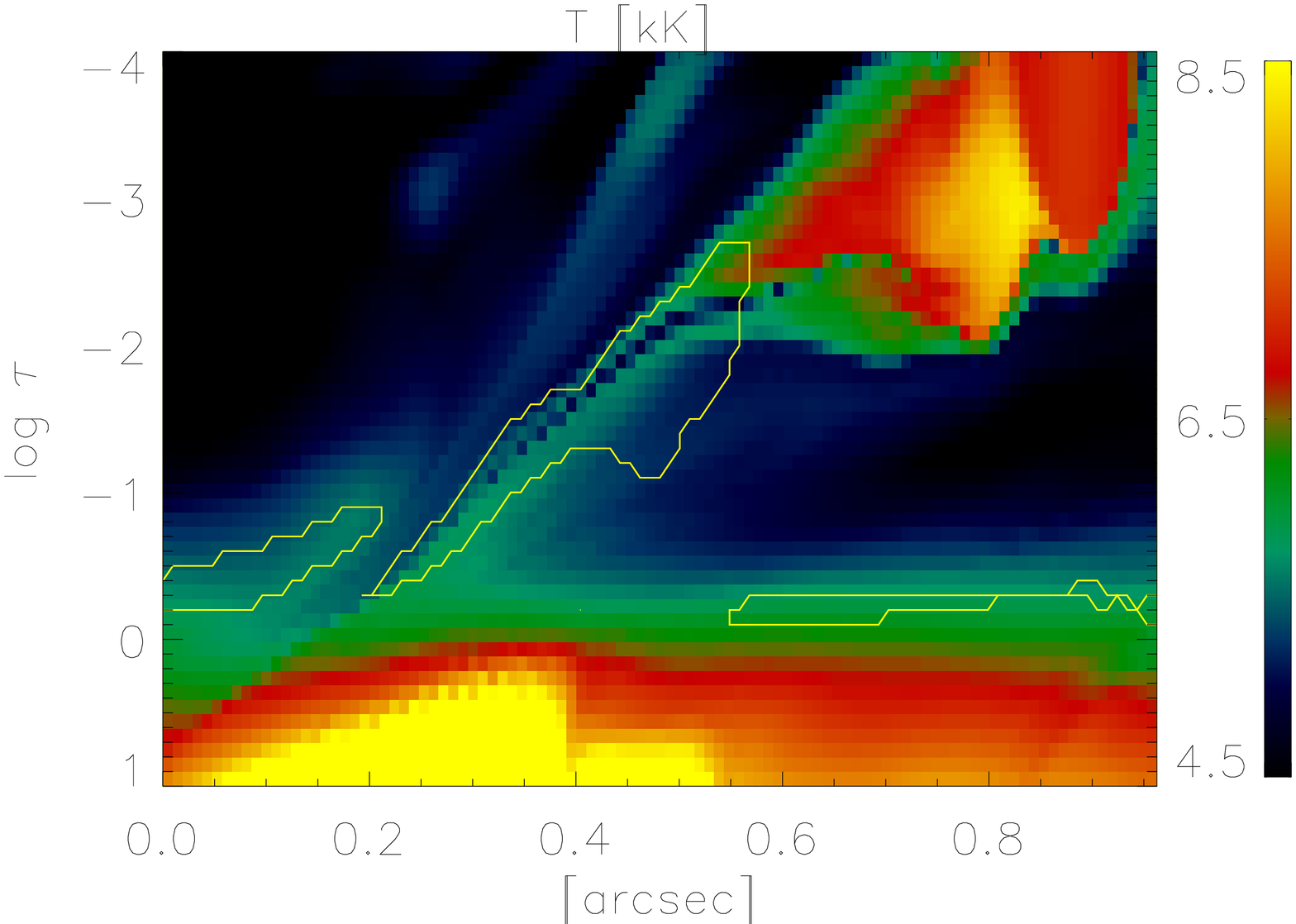}
  \includegraphics[angle=90,width=0.32\linewidth]{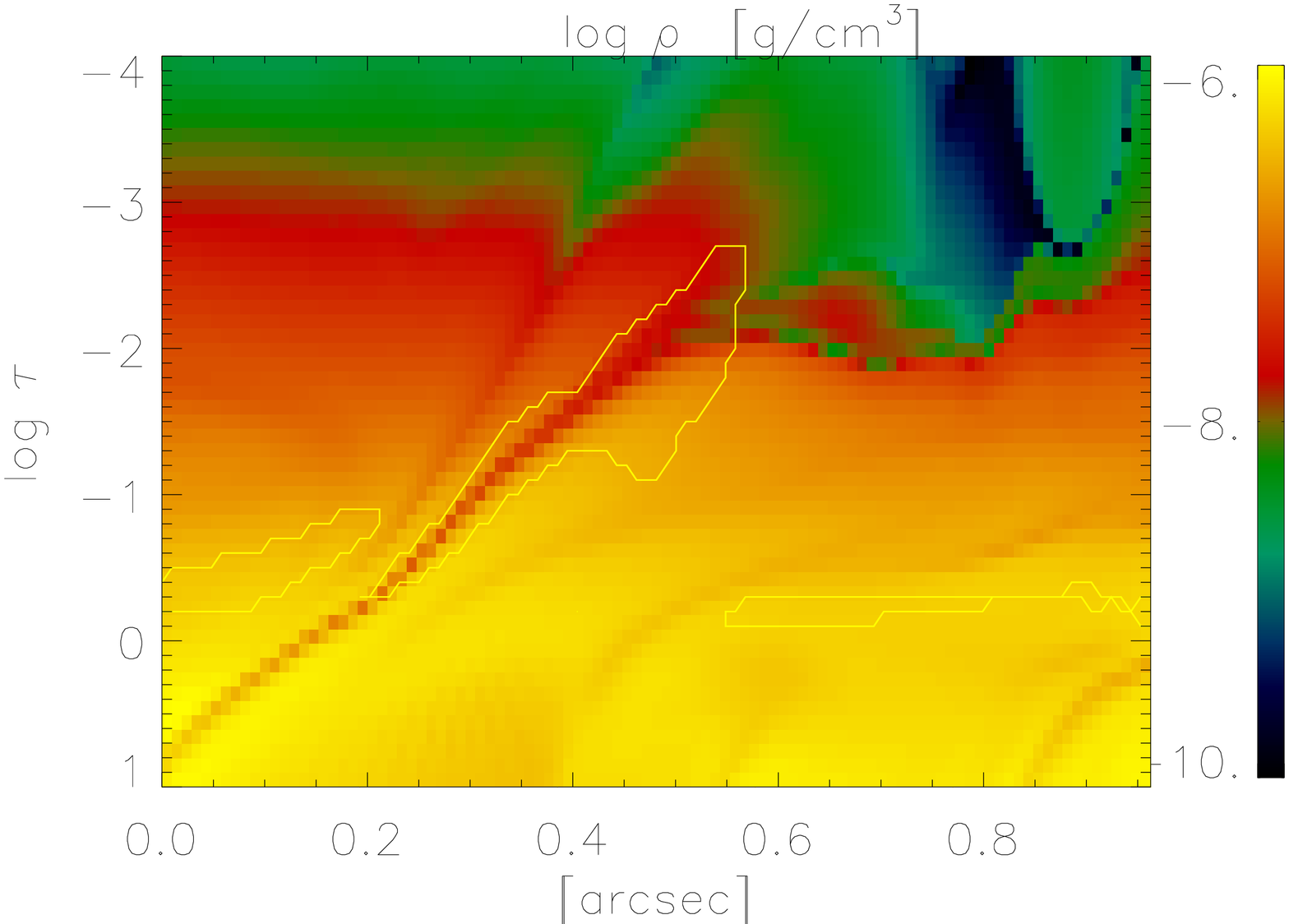}
   \includegraphics[angle=90,width=0.32\linewidth]{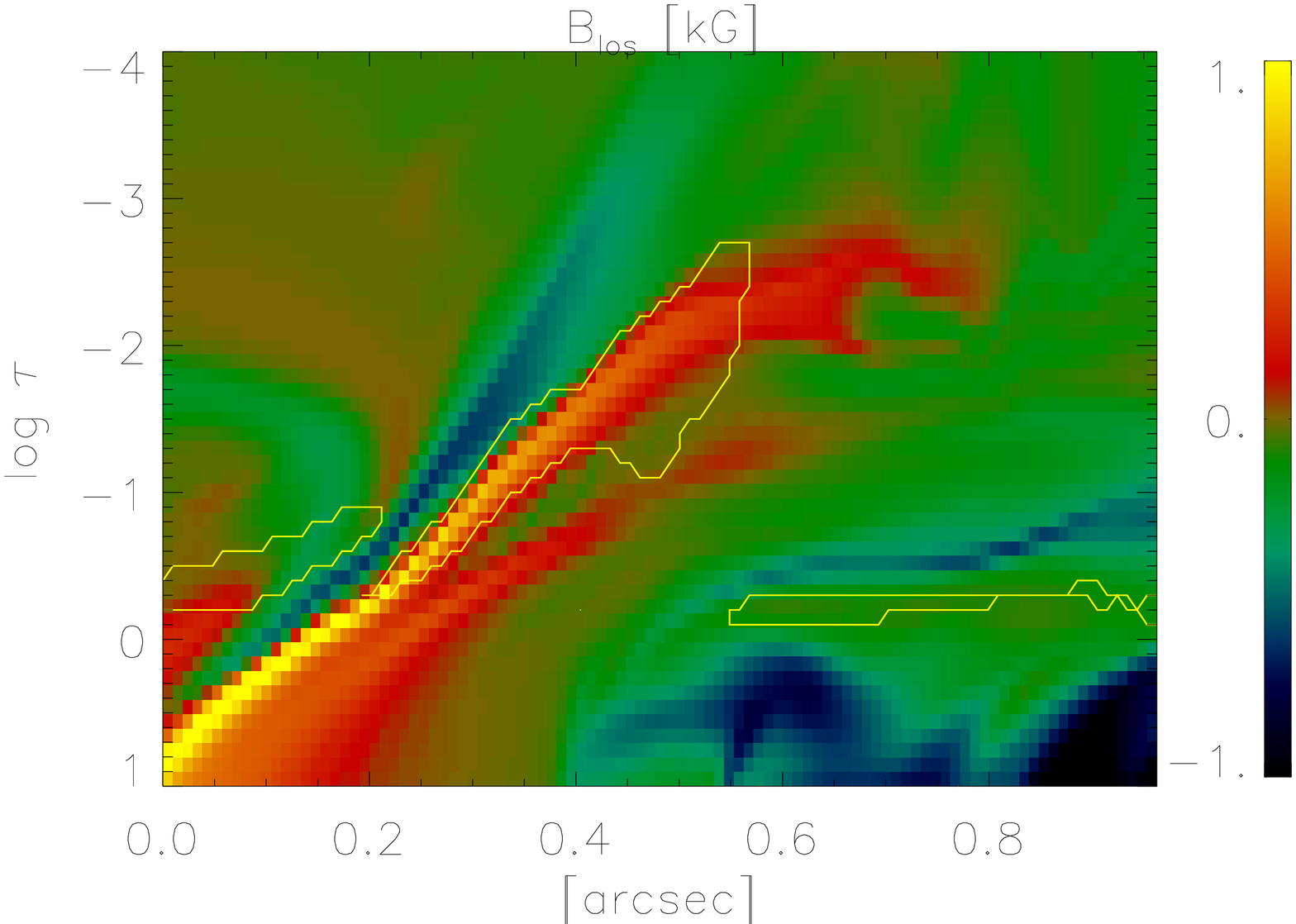} \\
  \includegraphics[angle=90,width=0.32\linewidth]{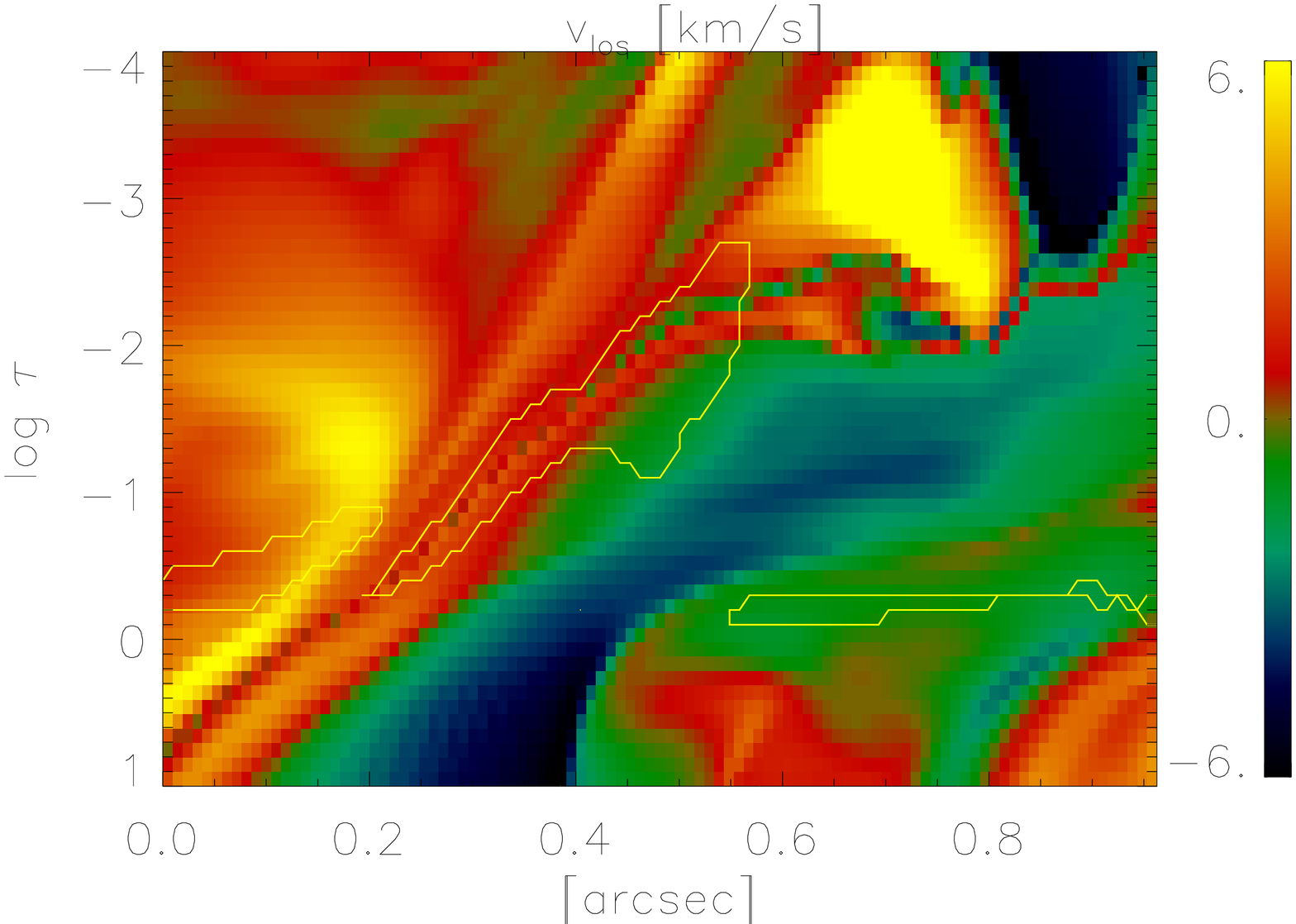} 
  \includegraphics[angle=90,width=0.32\linewidth]{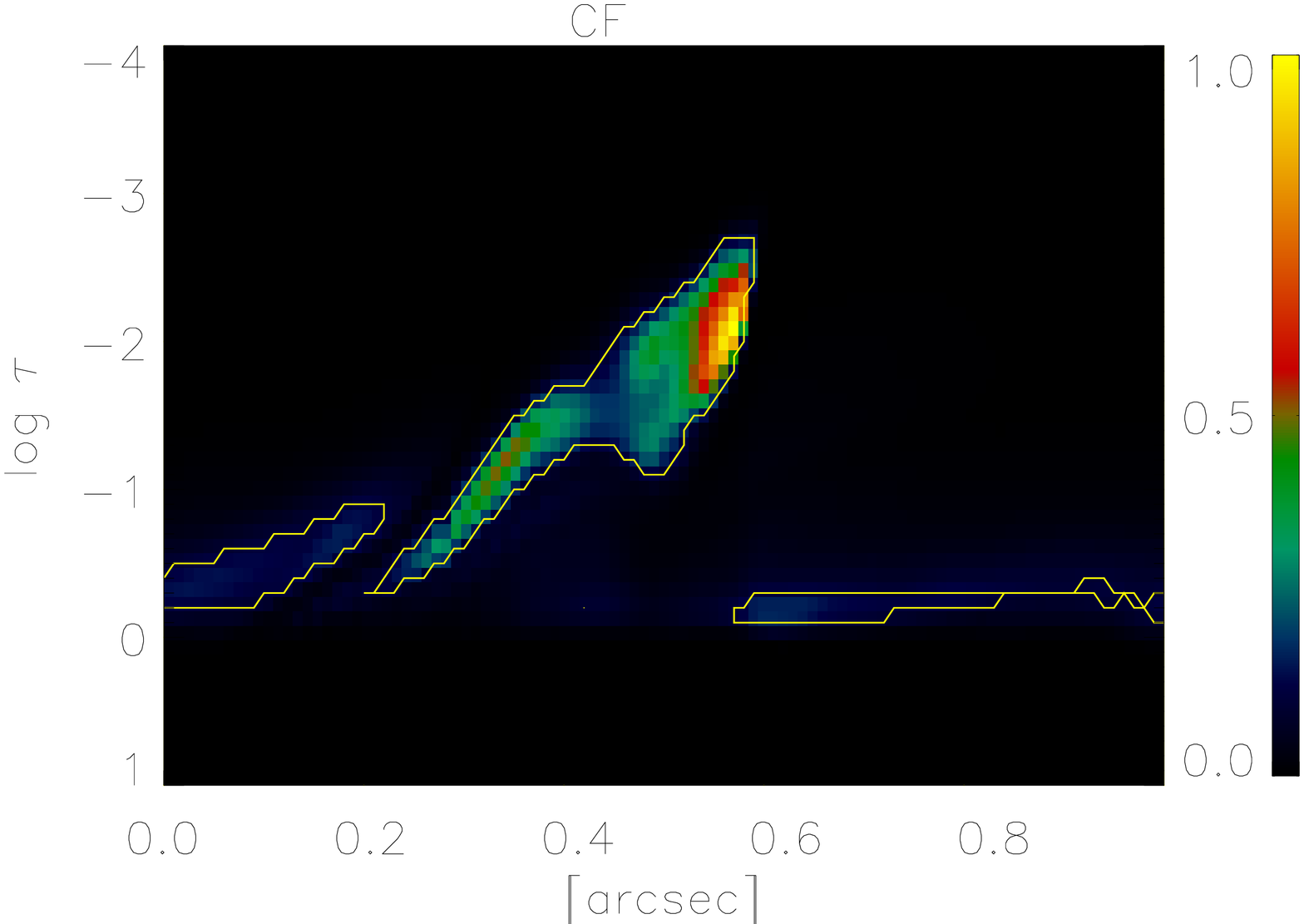} 
  \includegraphics[angle=90,width=0.32\linewidth]{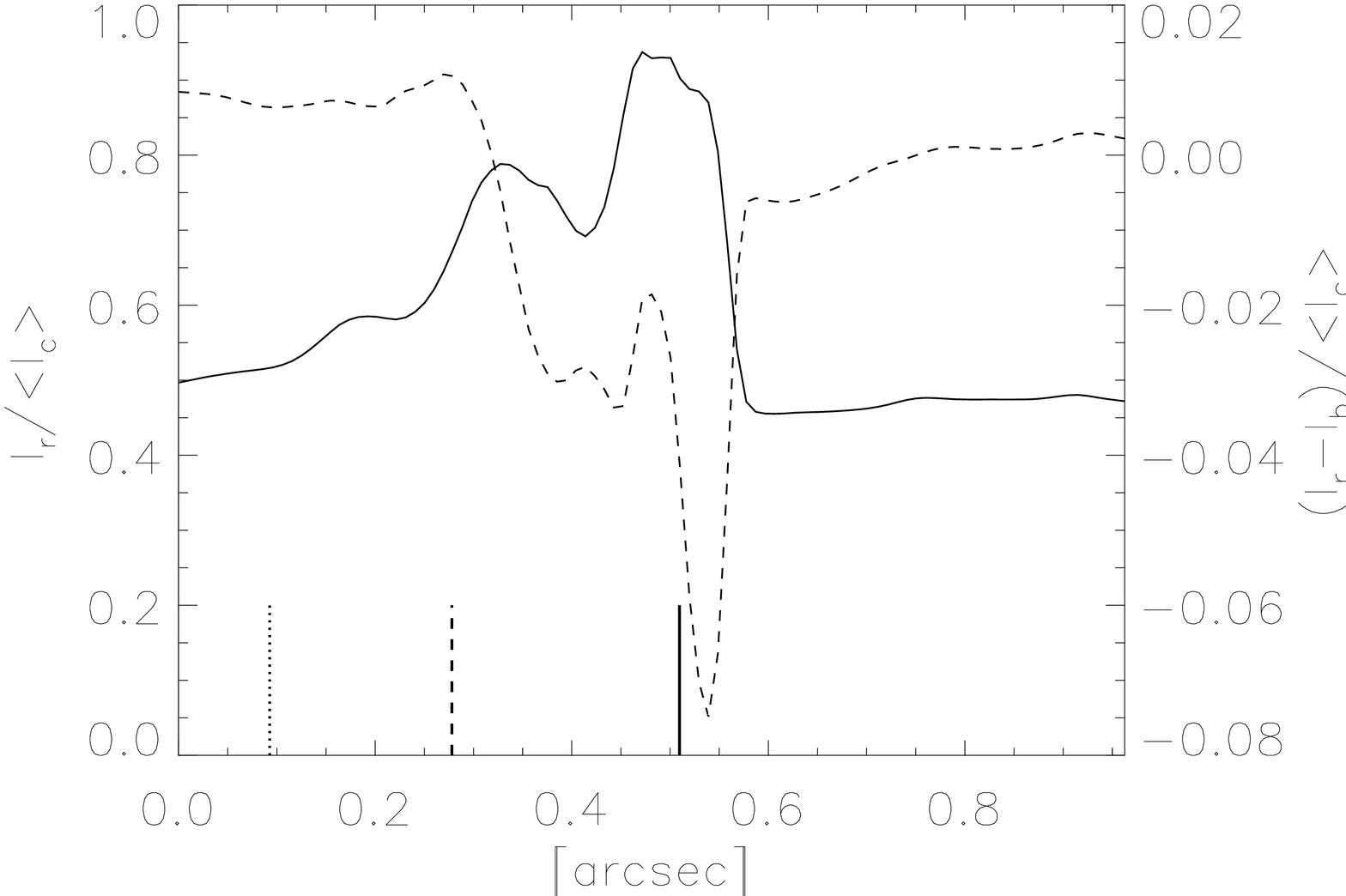}
  \caption[]{\label{fig:muram1-cuts-th49} %
  Vertical plane through the skewed simulation with viewing angle
  $\theta = 49^{o}$ and azimuthal angle $\phi = 90^{o}$,
  along the cut segment specified in the righthand column of
  Fig.~\ref{fig:muram1-cuts-map}.
  The format is the same as in Fig.~\ref{fig:muram1-cuts-th0}.  
  The $\tau$ scales along the $y$-axes are now continuum optical depth
  at 5000~\AA along the slanted lines of sight through the cut plane
  (which has the solar limb to the right).
  The $x$-axis scales are apparent distance over the solar surface
  including foreshortening.
  }
\end{figure*}


\section{MURaM simulations}
\label{sec:muram}
Two essentially different numerical experiments are performed with 
the 3D MHD MURaM code \citep{2005A&A...429..335V,Rempel:2009} to simulate the quiet Sun and active regions conditions. The runs are variants of those used in \cite{ja:jets}. Both ambipolar diffusion and Hall currents are taken into account in the induction equation, because these might be important in the photosphere for dynamics at very small scales, especially in
regions with large magnetic field gradients
\citep{Mark:ambi:2012}.  Non-grey radiative transfer is also included.

\subsection{Quiet Sun case}
\label{sec:qs}

The first run presents a case rather similar to the one of
\cite{Nelson2013}. 
The computational domain measured $6\times6\times1.68$~Mm with about
$700$~km above the $\tau_{5000}=1$ surface (continuum optical depth
at $\lambda = 5000$~\AA) and with a spatial resolution of $10$ and
$14$~km in the horizontal and vertical directions, respectively. The initial magnetic field at the simulation start was
that of a $2\times2$ checkerboard pattern of bipolar vertical
magnetic field with height-independent strength of $B = 200$~G.

When the simulation had evolved during approximately 15 minutes of
solar time, a horizontal flux sheet was inserted at a depth of
$300$~km below mean optical depth unity.
The field strength profile in the vertical direction across the sheet
was defined as a Gaussian with a $\mbox{FWHM}$ of $110$~km and a maximum
value $1$~kG.
We then let the simulation run for additional 30 minutes of solar time
saving snapshots at half-minute intervals.

The collected snapshot are used to synthesize emergent profiles of H$_{\alpha}$ and other lines
along every column. In order to facilitate direct comparison with the H$_{\alpha}$ observations of
\cite{Gregal:2013} 
the same spectral sampling and smearing of 66~m\AA\ is used as in these
SST observations. Synthetic images in the line wing were produced with the SPINOR code of \cite{Frutiger2000,Frutiger:etal:2000} 
and more detailed profile synthesis was peformed along columns with the RH
code of \cite{Uitenbroek2001}. 
Both codes produce qualitatively similar results, but minor
differences arise from different representations of the hydrogen atom
and/or different numerical schemes. In SPINOR, linear Stark broadening is added as prescribed by \cite{Rutten:2016}. Because of this, the focus in this paper is given only to SPINOR output.

Figure~\ref{fig:muram1-3d} shows the event
that produced the strongest brightening in the H$_{\alpha}$ wing in this
simulation run.
It corresponds to a sizeable temperature increase that occurs at the site of the
cancellation of two magnetic features of opposite polarities.
Figure~\ref{fig:muram1-3d} outlines this feature by color-coding the
temperature increase in the region where the largest Ohmic heating
occurs.
Figure~\ref{fig:muram1-evolution} displays the same feature as it
appears in synthetic H$_{\alpha}$ wing images, synthetic H$_{\alpha}$
Dopplergrams, and Fe~I~6301~\AA~magnetograms, viewed from aside along
the slanted $\mu = 0.66$ line of sight in the four azimuthal
quadrants.
The feature appears as an upright ``flame'' rooted at the surface
location where the cancelation occurs,
very similar to the morphology described in
\cite{Watanabe2011} 
and \cite{Gregal:2013}. 

The synthetic H$_{\alpha}$ Dopplergrams in the second column of
Fig.~\ref{fig:muram1-evolution} suggest downflow at the base and
upflow higher, at all azimuthal perspective.

The magnetograms in the third column of
Fig.~\ref{fig:muram1-evolution} show that the small patch of positive
polarity that is present at the base of the green feature in
Fig.~\ref{fig:muram1-3d} is not clearly visible in the slanted view. 
The region around the EB-like feature actually appears unipolar when
observed from most angles.
This hiding implies that even when no bipolarity is observed, there
may actually be enough that cancellation produces EB-like phenomena.

Figure~\ref{fig:muram1-cuts-map} shows how this EB-like feature would
look at disk center with the spatial resolution of the simulation (top
row) and at the SST resolution (middle row).
Each panel is grey-scaled independently.
At the SST resolution the feature's brightness contrast is reduced
considerably.
At disk center it would not be very distinct from the brightest
network field concentrations, but in limb-ward viewing its flame-like
appearance would still stand out.
The magnetograms at the simulation resolution in the bottom panels
illustrate how small-scale intricate the field topography is which 
causes this feature.

Since a numerical simulation, as distinct from observations, permits one to
diagnose what actually happens also below the surface vertical
cuts through the EB-like feature in the $t = 62$~s snapshot is shown in
Fig.~\ref{fig:muram1-cuts-th0}.
Their location is specified in the lefthand top and bottom panels of
Fig.~\ref{fig:muram1-cuts-map}.
The third panel shows how this cut samples the two opposite-polarity
magneric concentrations of which the partial cancelation caused the
EB-like feature. Higher up these magnetic features happen to lie out of the selected
cut plane, so that especially the orange feature in the third panel of
Fig.~\ref{fig:muram1-cuts-th0} shows an abrupt top due to end of
sampling.


The first panel of Fig.~\ref{fig:muram1-cuts-th0} shows a substantial
temperature increase in and especially just above the sampled part of
the righthand magnetic concentration, up to $T = 8500$~K at $h
\approx 300$~km but starting already at the surface (with $h = 0$
defined by $\tau_{5000} = 1$).

The second panel of Fig.~\ref{fig:muram1-cuts-th0} shows 
the presence of the two magnetic concentrations as relatively
tenuous ``fluxtubes''.  However, a local 
density increase is present above the second one at the height
where the EB-like feature reaches its highest temperature.

The simultaneous increase of temperature and density in the EB-like
feature suggest that reconnection has occurred and caused the
apparent partial cancelation of the white patch against the black one
in Fig.~\ref{fig:muram1-3d}. 
These opposite-polarity patches of strong field are closely adjacent.
The corresponding field lines in Fig.~\ref{fig:muram1-3d} suggest
closed loops between them, similar to the case presented in \cite{2010ApJ...720..233C}. The apparent cancellation and heating at this site suggests reconnection between the legs of a $\cap$-loop as they  approach or they are being pushed towards each other.


The second row of Fig.~\ref{fig:muram1-cuts-th0} displays the same
quantities, but against mean radial optical depth $\tau_{5000}$ for
the continuum at $\lambda = 5000$~\AA\ and adding contours that
outline the contribution function to the radially emergent intensity
in the H$_{\alpha}$ wing, as defined in the center panel of the bottom
row.
The first panel of the second row shows that the hot top of the
EB-like feature sampled by the cut plane is optically thick in the
H$_{\alpha}$ wing, so that its lower parts are shielded from view.
The lines of sight to the left of it sample a cool intergranular lane
while the lines of sight to the right sample the edge of large hot
granule.
In both adjacent areas the H$_{\alpha}$ wing originates much deeper, close
to $\tau_{5000}= 1$.
This is expected since the cool upper photosphere of typical standard
models contains no H$_{\alpha}$ opacity due to the low Boltzmann
excitation of its lower level, whereas
the higher temperature in the EB-like feature enhances the H$_{\alpha}$
opacity substantially.
Note also that the relatively large contrast that magnetic
concentrations obtain in the blue H$_{\alpha}$ wing thanks to its deep
formation
\citep{Leenaarts:etal:2006} 
is indeed present in Fig.~\ref{fig:muram1-cuts-map}.

The density in the second panel of the second row shows a striking dip
at the location of the density excess in the first row.
This difference results from the transition from geometrical height to
a $\tau_{5000}$ scale and implies larger continuum opacity in the
EB-like feature. 

In the third panel of the second row the H${\alpha}$ -wing formation
contour sits just above the abrupt end of sampling the righthand
magnetic concentration.

The first panel of the third row of Fig.~\ref{fig:muram1-cuts-th0}
shows that the EB-like feature has a considerable downflow, and that
an upflow lies above it and a bit to the side.
The second panel defines the outline of the contribution to the
emergent H$_{\alpha}$ -wing intensity used in the other panels. 
The third panel shows this emergent intensity (solid) across the cut,
and also the subtractive Dopplergram signal of which the deep dip
corresponds well to the large downdraft seen at left.

This diagnosis indicates that in disk-center viewing along the radial
direction one sees the top of an EB only and that that shields what
lies underneath.
In addition, the third panel of Fig.~\ref{fig:muram1-cuts-map}
demonstrates that even at the SST resolution such a feature may not be
easily distinguished from normal network concentrations.

Figure~\ref{fig:muram1-cuts-th49} is similar
to Fig.~\ref{fig:muram1-cuts-th0} and shares its format, but diagnoses
the skewed cube along the cut defined in the righthand column of
Fig.~\ref{fig:muram1-cuts-map}.
It lies along the direction to the limb (to the right) which the
feature would follow if it was vertical and straight (which it is
neither, but fairly close).
The slanted view now permits also sampling the lower parts of the
EB-like feature.
It results in extended elongated formation envelopes that reach down
to the surface.
This deeper-down sampling from aside indeed makes the bright
corresponding feature in the righthand column of
Fig.~\ref{fig:muram1-cuts-map}. Fig.~\ref{fig:muram1-cuts-map} indeed
display ``'flame'' morphology as in the SST observations. The H$_{\alpha}$-wing outlines the hot region and increases as the temperature increases, as shown in the bottom right panel of Fig.~\ref{fig:muram1-cuts-th49}. 

\begin{figure}
 \includegraphics[angle=90,width=0.99\linewidth]{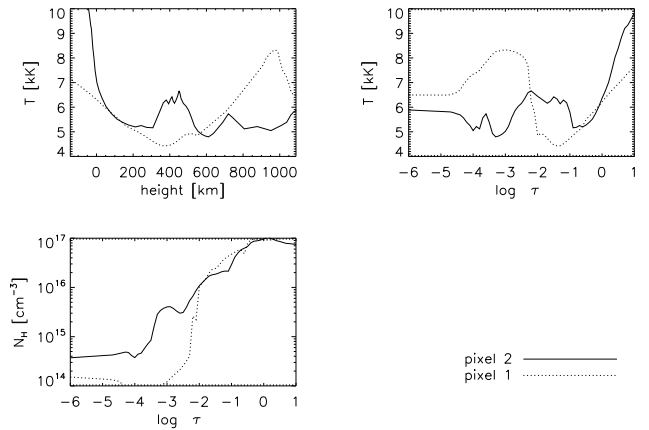} 
   \caption{Comparison of the atmospheric parameters at two locations: the middle (pixel 1) and top (pixel 2) of the feature. Top panels show temeprature as a function of geometrical height (left) and optical depth (right). Bottom left panel shows total hydrogen density along the line of sight as a function of optical depth.}
\label{pix_com}
\end{figure}

Finally, two positions along the features are chosen for quantitative comparison with 1D models. Fig.~\ref{pix_com} shows two examples that sample the EB-like feature at its middle and its top where the maximum temperature increase occurs. The height profiles show bumps where the temperature increases by 1500 to 3000 K. This agrees with prediction from one-dimensional modelling where the observed H${\alpha}$  wings are fit by introducing ad-hoc perturbations of a static 1D standard model of the solar atmosphere. Furthermore, pixel 2 shows that, in the slanted view representation, the top of the flame reaches 1 Mm along the line of sight, similar to what some 1D EB models proposed. However, the arising temperature of 8000~K combined with the total hydrogen density of the order of $10^{14}-10^{15}$~cm$^{-3}$ comes up short in comparison with constrains given by \cite{Rutten:2016} that could insure the EB's visibility in wide range of observables.

\subsection{Active region case}
\label{sec:muram2}

\begin{figure*}
  \centering 
   \includegraphics[width=0.99\linewidth,trim=0cm 0cm 0cm 0cm,clip=true]{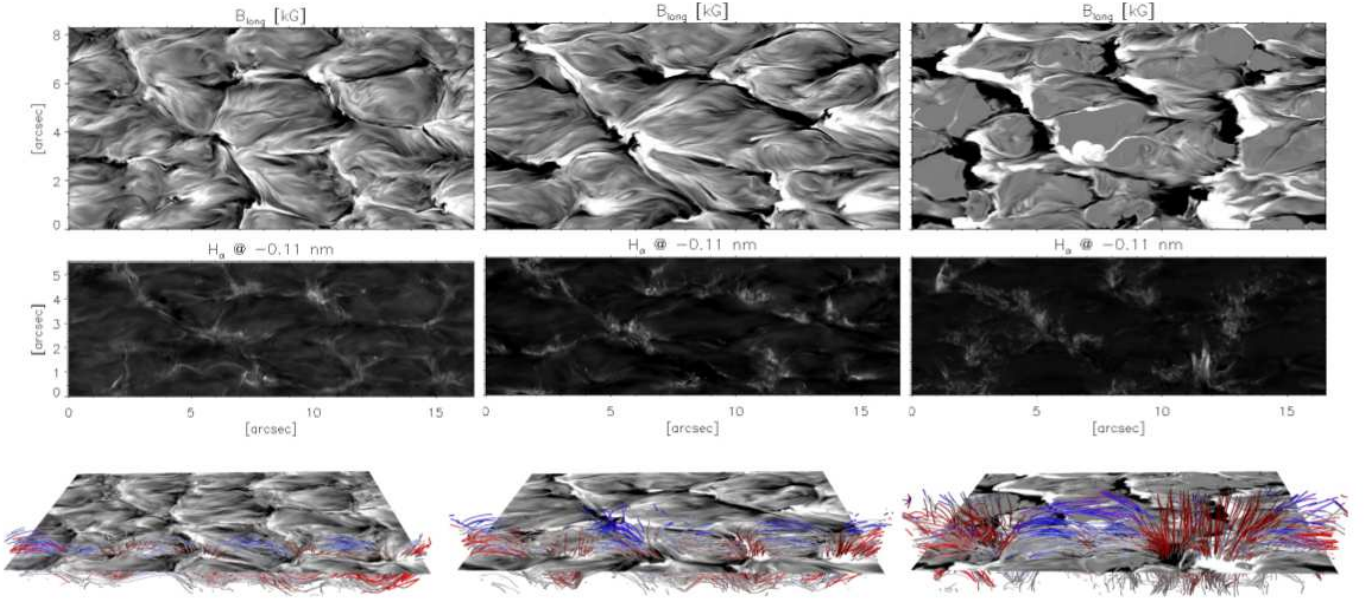}
    \caption{Vertical component of magnetic field (top row), H$_{\alpha}$ wing intensity (middle row) and field topology (bottom row) at t$=112, 197$ and $549$~s from left to right. The color coding of the field lines in the bottom panels corresponds to the vertical velocity with upflows being blue. Horizontal planes show vertical component of magnetic field at the surface. The magnetograms are cropped at $\pm1 $~kG and normalized H$_{\alpha}$ wing intensity at 0.7 to 4.5.}
\label{em_5000}
\end{figure*}

\begin{figure}
  \centering 
   \includegraphics[width=0.9\linewidth,trim=1.5cm 5cm 1.5cm 5.5cm,clip=true ]{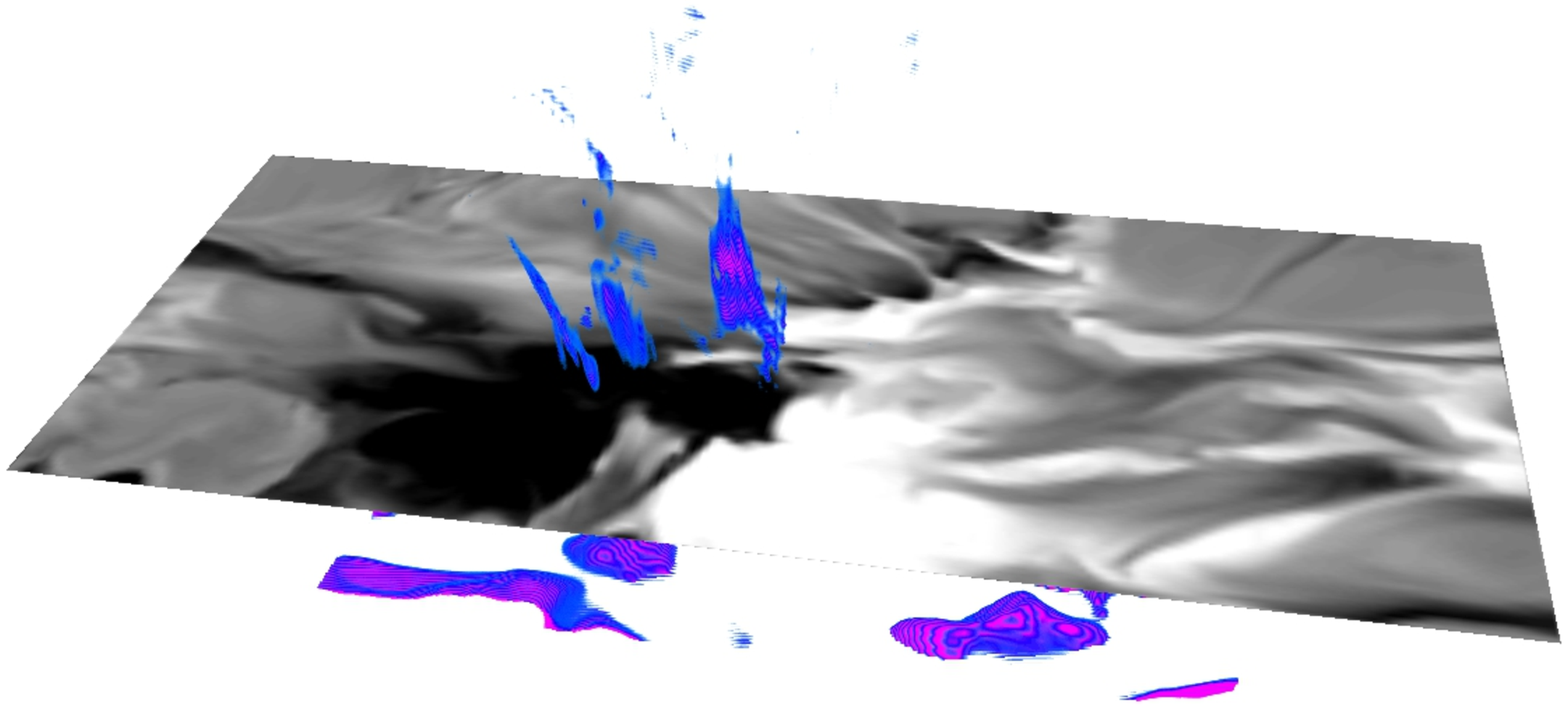}  
    \includegraphics[width=0.9\linewidth,trim=1.5cm 5cm 1.5cm 5.5cm,clip=true]{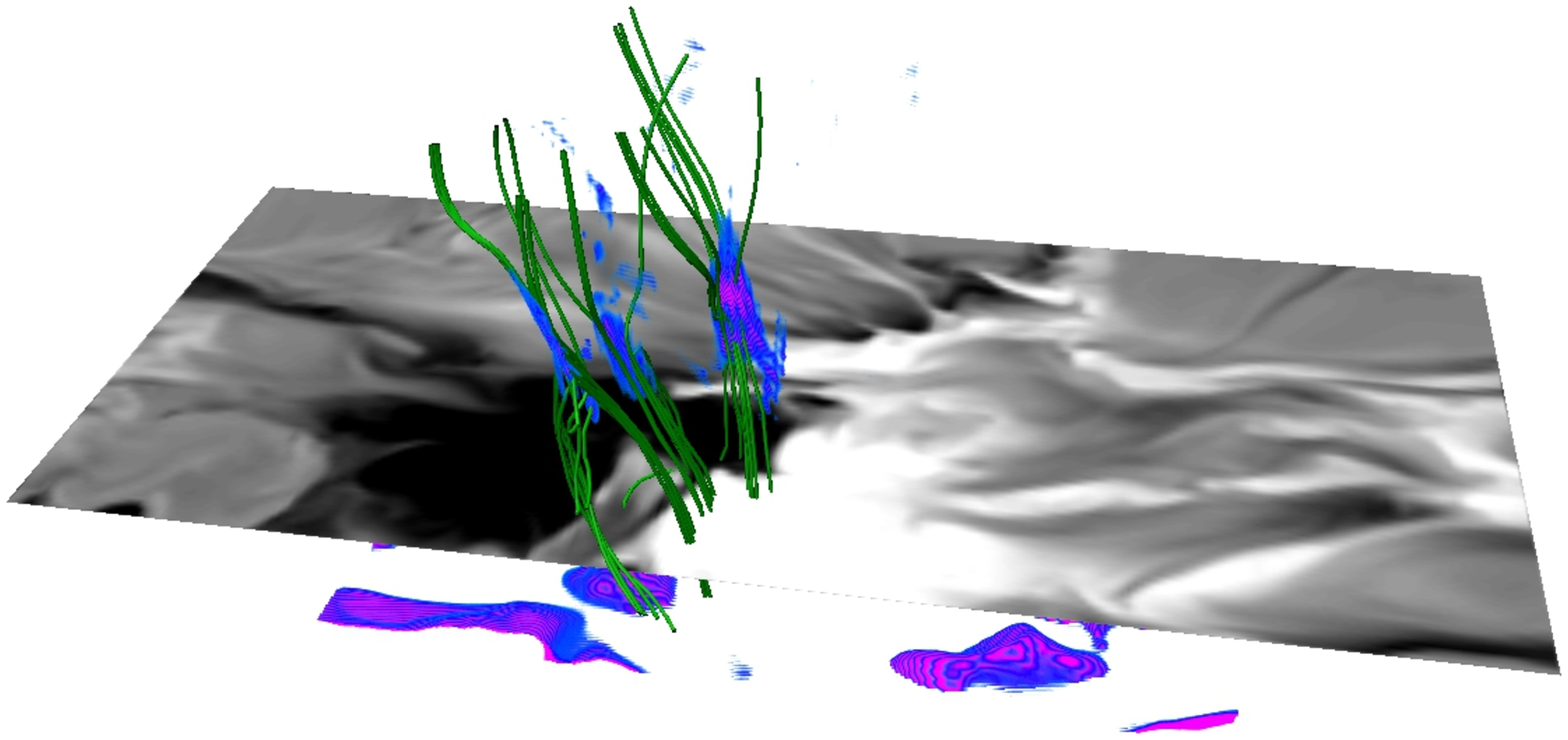} 
    \includegraphics[width=0.9\linewidth,trim=1.5cm 7cm 1.5cm 3.5cm,clip=true]{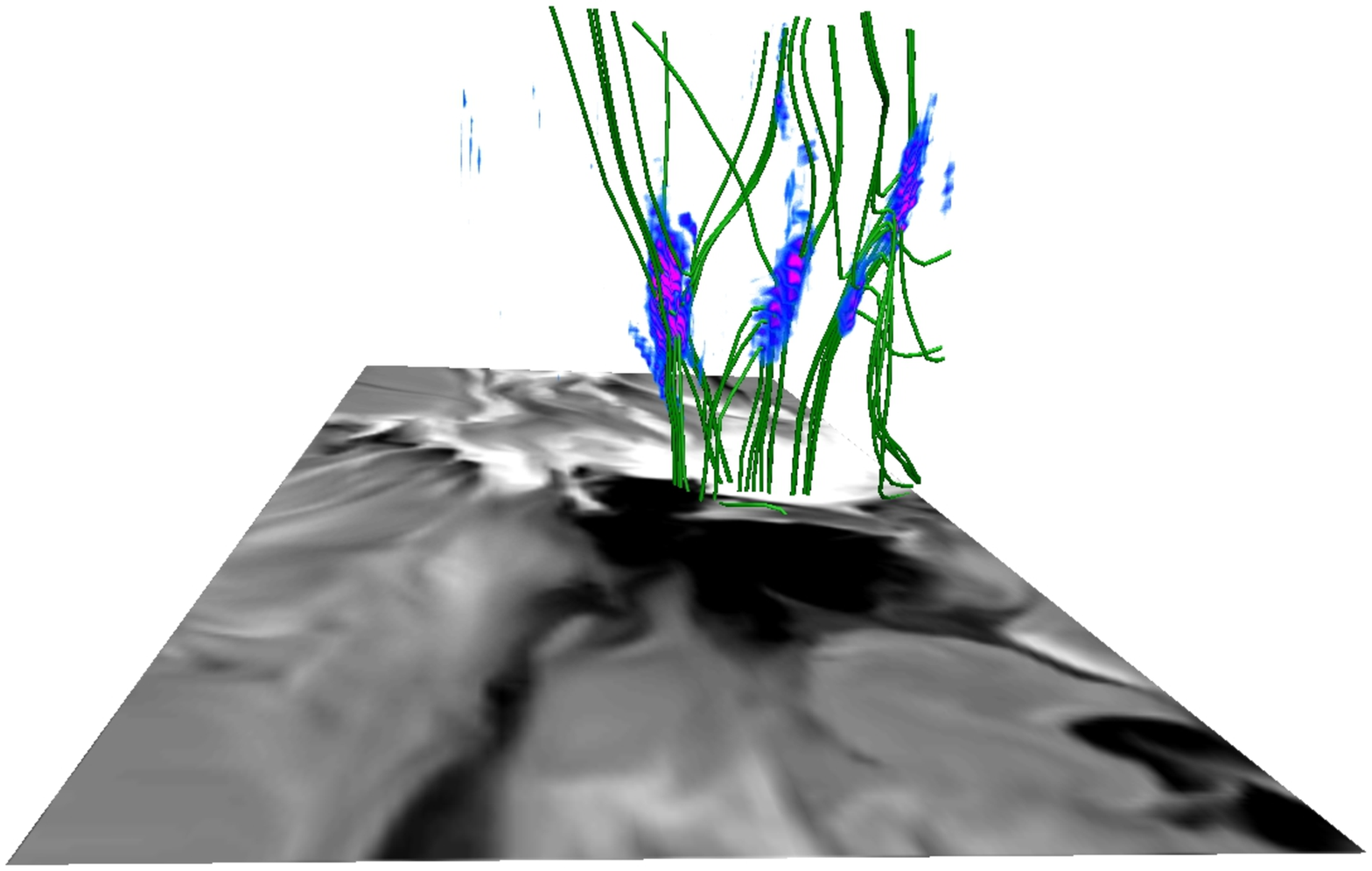}   
    \caption{Zoom in at a small part of the simulation domain at $t=549$~s at [11\arcsec,1\arcsec]. Horizontal panels show vertical component of magnetic field at the surface. Blue/purple rendering depicts the temperature in the range of 7500-9000~K and green lines show magnetic field configuration at those locations. Bottom panel shows the same region as the upper two but from a different viewing angle.}
\label{fork_3d}
\end{figure}

To simulate an active region case, this time the flux sheet is introduced in a purely hydrodynamic setup. The computational domain in this run is $12\times6\times3.5$~Mm with about $2$~Mm above the $\tau_{5000}=1$ surface and the same spatial resolution as in the quiet Sun run. The width of the flux sheet and location where it is introduced are the same as in the quiet Sun run, only in this case maximum field strength is 5000~G instead of 200~G. This gives maximal signed flux density of around 400~G, which roughly agrees with values detected in the active regions. The flux sheet gets undulated due to convection in such a way that crests/troughs are formed where convective uplows/downflows are present. Fig.~\ref{em_5000} shows vertical field, H$\alpha$ wings intensity and the field topology at three moments during the emergence. At the first moment, the top of the flux sheet reached the convectively stable layer, while the H$\alpha$-wing intensity map shows only weak brightenings at location where reconnection takes place just at the surface. First flame-like features appear few minutes after when the emerged loops reach a few hundred km above the surface. The third instance shows the phase when the longest and brightest features appear. At this point the whole flux sheet already emerged and the non-perturbed granular pattern is restored. Similarly as in the quiet Sun case, the most prominent EB-like features appear at the places where concentrated features of opposite polarities rush towards each other as the footpoints of the emerging loops travel horizontally.

In the active regions case, the morphology of EB-like features takes more complex forms than in the quiet Sun case, as a natural result of more intricate field topology. Fig.~\ref{fork_3d} shows in detail how the 'fork'-shaped feature visible at [11\arcsec,1\arcsec] in the left-most middle panel of Fig.~\ref{em_5000} is generated. The temperature increase and the field configuration are similar to the quiet Sun case, only here it happens at three locations simultaneously. The largest jumps in temperatures occur near the temperature minimum where the temperature reaches 9000~K. The field topology in all three location is similar. It shows the bottom of the current sheets and reconnected  
$\cap$-loops. 

\begin{figure}
  \centering 
   \includegraphics[angle=90,width=0.9\linewidth,trim=2cm 0cm 8.8cm 1cm,clip=true]{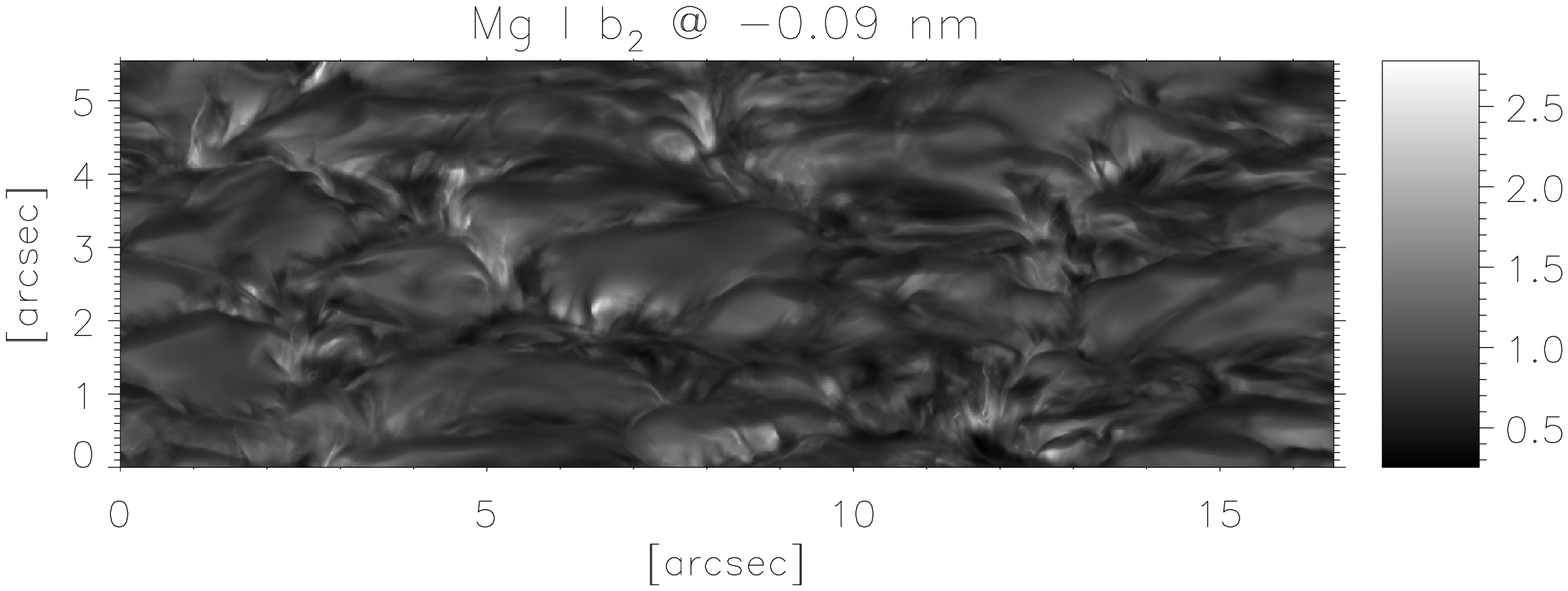} 
  \includegraphics[angle=90,width=0.9\linewidth,trim=2cm 0cm 8.8cm 1cm,clip=true]{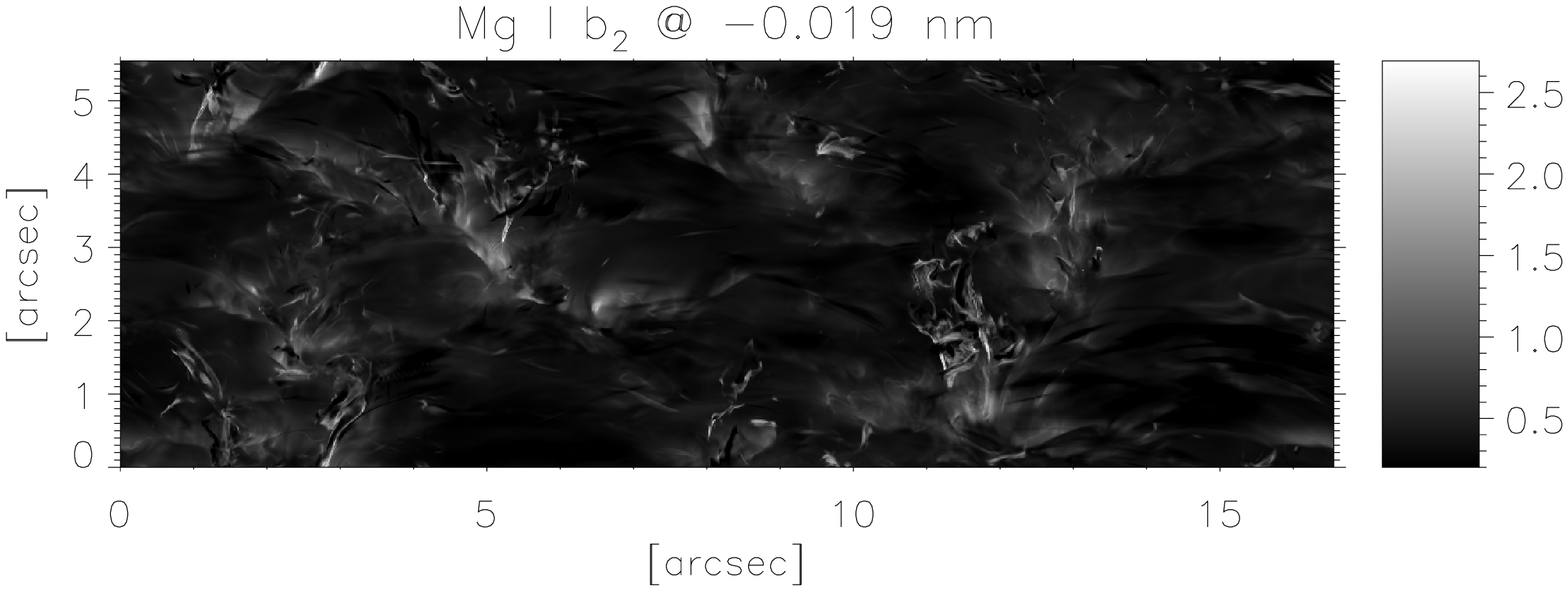} \\
  \includegraphics[angle=90,width=0.9\linewidth,trim=2cm 0cm 8.8cm 1cm,clip=true]{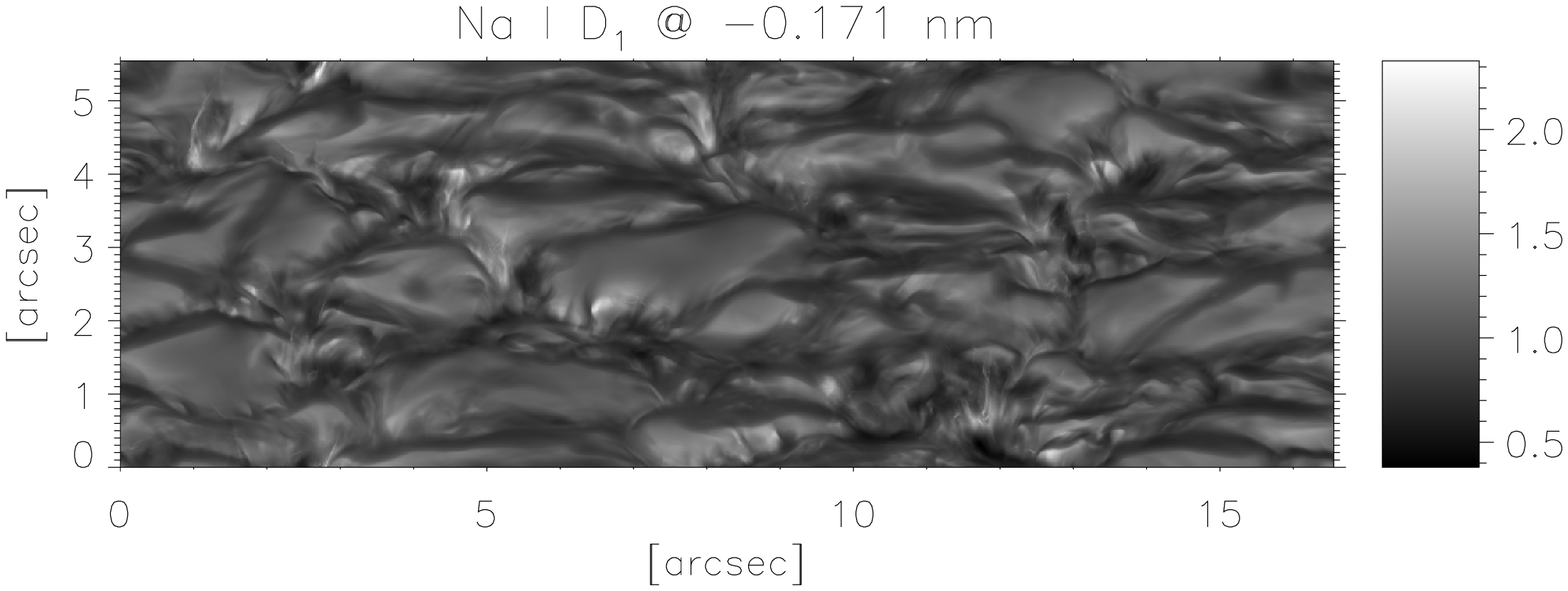} 
  \includegraphics[angle=90,width=0.9\linewidth,trim=0cm 0cm 8.8cm 1cm,clip=true]{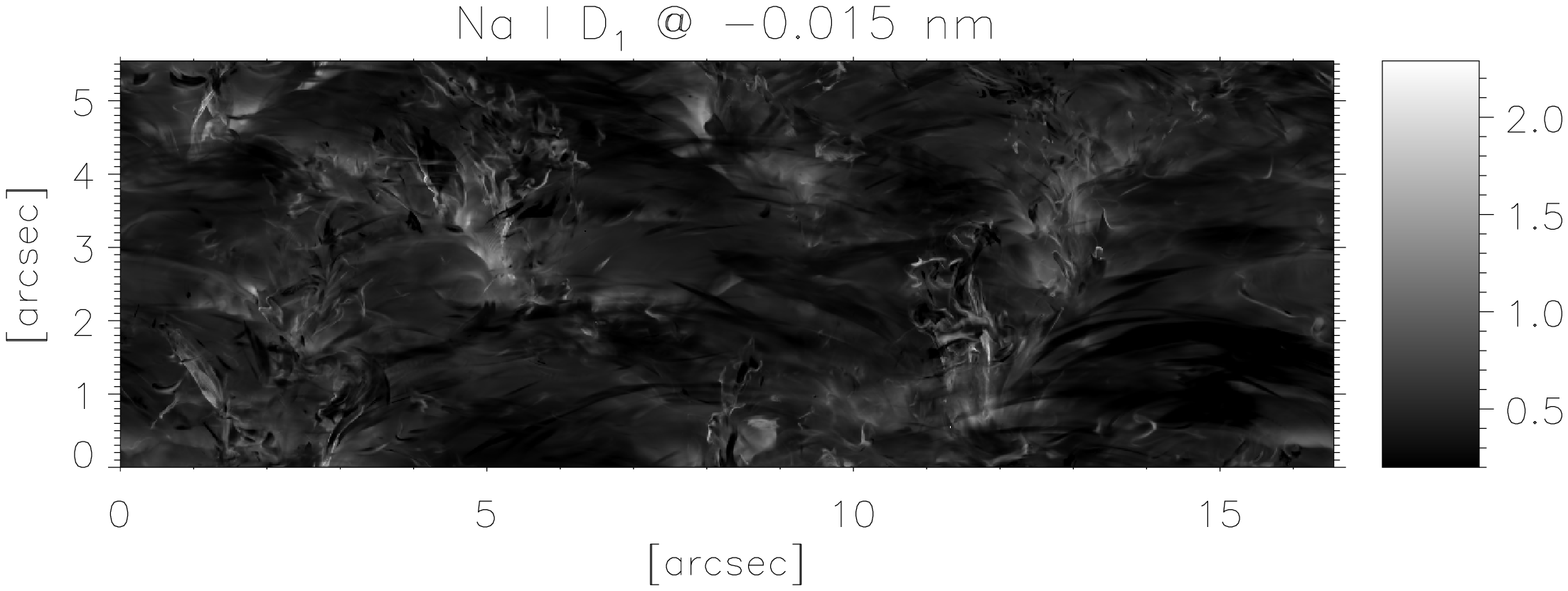}
    \caption{Synthetic filtergrams in the continuum and the wing of Na~I~D1 and Mg~I~b1 lines. The snapshot corresponds to the instant $t=549$~s shown in Fig.~\ref{em_5000}.}
\label{other_lines}
\end{figure}

Finally, Fig.~\ref{other_lines} shows how these features look in the continuum and the wings of Na~I~D1 and Mg~I~b1 lines. The chosen wavelength positions are the same as in observations obtained by \cite{Rutten2015}. Although observations how that no signatures of EB can be visible in these lines, synthetic filtegrams show brightenings at the same locations as the H$\alpha$ wings intensity image. Corresponding H$\alpha$ wings in Fig.~\ref{em_5000} shows three EB-like features: at [1\arcsec,5\arcsec],[8\arcsec,5\arcsec] and [11\arcsec,1\arcsec]. The first two are clearly noticeable also in  Na~I~D1 and Mg~I~b1 lines. The third, the most pronounced one in  H$\alpha$ wing can be traced also in Na~I~D1 and Mg~I~b1 images, but only at its foot point. The top and most of the 'fork'-like shape is missed by Na~I~D1 and Mg~I~b1 lines. Comparison of the temperatures in these locations reveals that although the temperature increases at approximately the same heights, in the first two cases it stays below 8000~K. 
In the third case, it is by more than 1000~K higher which comes into the temperature range where Mg I and Na I population drops rapidly \citep{Rutten:2016}.

\section{Conclusion}  \label{sec:conclusion}

EB-like features are here simulated by using realistic 3D MHD code MURaM. Two different cases are presented: the quiet Sun case that resembles the QSEBs conditions and the active region case that tries to reproduce the serpentine like flux emergence. Although the simulation domain reaches 2 Mm above the solar surface in the active region case, the analysis is limited to only the lower atmospheric layers which are properly represented with this code. The results reveal that several characteristics of EB are replicated by the simulations.

Firstly, EB-like brightenings coincide with hot and dense locations, in agreement with predictions of 1D and 2D modellings. In the simulations, the hot clouds appear at or just bellow the temperature minimum which supports the claim that EBs are photospheric phenomena. The temperatures at these location are, in some cases, higher than 9000~K which is enough to reproduce the weak end of EB spectrum. This however does not exclude the possibility that higher temperature could be produced higher up the developed current sheets. The simulations show though that the wings of H$\alpha$ sample always the low atmospheric layers in these runs. In the hottest cases, the features visible in H$\alpha$ wings will not be present in Na~I~D1 and Mg~I~b1  filtergrams which is in agreement with observations.

Secondly, the simulated EB-like phenomena has the observed flame-like morphology with the base rooted in intergranular lanes. As shown in this study, the slanted view reveals the lower parts of the EB-like feature which results in extended elongated formation envelopes that reach down
to the surface. The simulated features assume the direction perpendicular to the 'limb'. In some cases, though, depending on the viewing angle, the form can be more complex and the flame can be observed at an angle. These complex forms are more often found in the active region case.

Thirdly, simulated EB features are caused by reconnection apparent as the cancellation of strong-field patches of opposite polarity
that move together. At the layers sampled by  H$\alpha$-wings, the magnetic field topology seem to be always similar. The field lines there trace the base of the current sheet and the reconnected $\cap$-loops. This is the same in both the quiet Sun and the active region cases. Although the opposite polarities are always present, this however might not be always detected in observations due to projection effects. The detection is further impaired with decrease in spectral and spatial resolution and poor polarimetric sensitivity. This suggests that shearing magnetic field hypothesis \citep{2008ApJ...684..736W,Gregal:2013} might not be needed to explain the unipolar EB cases \citep{Qiu:2000,Georgoulis:2002,2008ApJ...684..736W,Hashimoto2010}.

Finally, another similarity between two runs is that the EB-like features do not appear in all regions where opposite polarities cancel out. They seem to appear in cases where opposite polarities approach each other at a reasonable high speed. As we showed in \cite{ja:sunrise2}, EBs form and last as long as the surface flows driving them together persist and can reappear or increase in brightness during re-emergence. This agrees with observations which also show that the EBs appear in the regions where the surface flows are the strongest \citep{Gregal:2013,Reid2016,Nelson2016}, due to large scale emergence or the moat flows. Based on this, one could also argue that QSEBs are triggered during a rapid  emergence on smaller scales.

%

\begin{acknowledgements}
This work has benefited from the discussions at the meeting 'Solar UV bursts – a new insight to magnetic reconnection' at the International Space Science Institute (ISSI) in Bern. NASA's Astrophysics Data System (ADS) and the 3D visualisation tool Vapor  were extensively used in this study. 
\end{acknowledgements}


\end{document}